\documentclass[manuscript,screen,table, xcdraw]{acmart}
\AtBeginDocument{%
  }

\setcopyright{acmlicensed}
\copyrightyear{2025}
\acmYear{2025}
\acmDOI{XXXXXXX.XXXXXXX}

\setcopyright{none}
\renewcommand\footnotetextcopyrightpermission[1]{}
\usepackage{tikz}
\usepackage{amsmath}

\usepackage{pifont}
\newcommand{\cmark}{\ding{51}}%
\usepackage{filecontents}
\usepackage{xspace}
\usepackage{multirow}
\usepackage{float}
\usepackage{caption, subcaption, svg, graphicx}

\begin{document}

\definecolor{BrickRed}{HTML}{B6321C}
\definecolor{RoyalBlue}{HTML}{4169E1}

\newcommand{\sysname}{VectorCDC\xspace}

\newcommand{\samer}[1]{\textcolor{green}{\textbf{[Samer: #1]}}}
\newcommand{\harsha}[1]{\textcolor{BrickRed}{\textbf{Harsha: #1}}}
\newcommand{\abed}[1]{\textcolor{purple}{\textbf{Abed: #1}}}

\newcommand{\newtext}[1]{\textcolor{black}{#1}}
\newcommand{\oldtext}[1]{}

\title{Accelerating Data Chunking in Deduplication Systems using Vector Instructions}

\author{Sreeharsha Udayashankar}
\email{s2udayas@uwaterloo.ca}
\affiliation{%
  \institution{University of Waterloo}
  \city{Waterloo}
  \country{Canada}
}

\author{Abdelrahman Baba}
\email{ababa@uwaterloo.ca}
\affiliation{%
  \institution{University of Waterloo}
  \city{Waterloo}
  \country{Canada}
}

\author{Samer Al-Kiswany}
\email{alkiswany@uwaterloo.ca}
\affiliation{%
  \institution{University of Waterloo}
  \city{Waterloo}
  \country{Canada}
}
\renewcommand{\shortauthors}{Udayashankar et al.}


\begin{abstract}

Content-defined Chunking (CDC) algorithms dictate the overall space savings that deduplication systems achieve. However, due to their need to scan each file in its entirety, they are slow and often the main performance bottleneck within data deduplication. We present \sysname, a method to accelerate hashless CDC algorithms using vector CPU instructions, such as SSE / AVX. \newtext{We analyzed the state-of-the-art chunking algorithms and discovered that hashless algorithms primarily use two data processing patterns to identify chunk boundaries: \textit{Extreme Byte Searches} and \textit{Range Scans}. \sysname presents a vector-friendly approach to accelerate these two patterns. Using \sysname, we accelerated three state-of-the-art hashless chunking algorithms: RAM, AE, and MAXP.} Our evaluation shows that \sysname is effective on Intel, AMD, ARM, and IBM CPUs, achieving $8.35\times$--$26.2\times$ higher throughput than existing vector-accelerated algorithms, \newtext{and $15.3\times$--$207.2\times$ higher throughput than existing unaccelerated algorithms}. \sysname achieves this without affecting the deduplication space savings.

\end{abstract}




\maketitle

\section{Introduction}

The amount of data generated and stored on the Internet is growing at an exponential rate \cite{statista2024}, and is expected to exceed 180 zettabytes per year in 2025. Storage capacity alone is not well positioned to handle this data influx, with the total installed storage capacity in 2020 only being 6.7 zettabytes \cite{statista2024}. Cloud storage providers instead support this data growth using alternatives such as novel storage paradigms \cite{software_defined_storage, raid}, distributed file systems \cite{hadoop, ceph} and caches \cite{memcached, facebook_tao}, mechanisms such as data deduplication \cite{dedup_intro, comprehensive_dedup}, alongside additional investments in data protection \cite{data_security}. 
 
Previous studies by Microsoft \cite{primary_dedup} and EMC \cite{backup_workload_charz} show that a large amount of redundancy exists in the data stored on the cloud, especially in file system backups \cite{primary_dedup, dedup_intro}, virtual machine backups \cite{backup_workload_charz, wandeltacomp} and shared documents \cite{documentduplication_2011}. The redundancy ratios in these workloads are significant, ranging from $50\%$ to $75\%$. Mechanisms such as data deduplication \cite{dedup_intro} and compression \cite{wandeltacomp} are used to conserve storage space by identifying redundant data, eliminating it, and minimizing its storage impact, thereby improving efficiency and reducing costs. 

Data deduplication consists of four phases \cite{comprehensive_dedup}; \textit{Data Chunking}, \textit{Chunk Fingerprinting}, \textit{Metadata Creation}, and \textit{Data Storage}. Data chunking and chunk fingerprinting are the most computationally intensive \cite{dedup_intro, dedupbench} of these. While chunk fingerprinting has received significant optimization attention, with faster hashing algorithms \cite{tunable_encrypted_dedup, scalable_incremental_checkpt} and GPUs \cite{gpu_dedup, storegpu}, data chunking optimizations have not kept up (\S\ref{sec:bg_dedup_bottlenecks}). 

In the data chunking phase, the incoming data is divided into small chunks, typically of size $1-64$ KB. Numerous data chunking algorithms exist in current literature \cite{ae, fastcdc, gear_hash, lbfs, ram, tttd} and can be broadly classified into hash-based and hashless algorithms \cite{dedupbench}. As chunking occurs whenever new data is uploaded, this phase is on the critical path, and directly impacts system performance.  

Previous efforts have explored accelerating chunking by using vector instructions.
SS-CDC \cite{sscdc} uses vector instructions to accelerate 
hash-based data chunking algorithms, such as Rabin-Karp chunking \cite{lbfs} and Gear-based chunking \cite{gear_hash}. Unfortunately, this approach only leads to modest improvements in chunking throughput, up to $3.13\times$, as shown in \S\ref{sec:bg_sscdc}. Parallelizing hash-based chunking using vector instructions is complicated because these algorithms use the rolling hash of a sliding window of bytes to detect boundaries, inherently creating a computational dependency between adjacent bytes. Consequently, SS-CDC processes different regions of the data in parallel using  
slow \texttt{scatter} / \texttt{gather} vector instructions, limiting its performance (\S\ref{sec:bg_sscdc}). 

We posit that hashless chunking algorithms are better candidates for vector acceleration. Although they achieve slightly lower space savings compared to their hash-based counterparts (\S\ref{sec:eval}), they are up to $2\times$ faster and use simple mathematical operations (e.g., finding a maximum value) that can be accelerated more efficiently using vector instructions. We analyzed state-of-the-art hashless algorithms to understand their design and identify opportunities to leverage vector instructions. We identified that all state-of-the-art hashless algorithms consist of two processing patterns. The first pattern involves finding local minima or maxima in a data region, which we call \textit{Extreme Byte Search}, and the second pattern involves scanning a range of bytes to find values that are greater or lesser than a target value, which we call \textit{Range Scan}. We found that, unlike rolling hash functions, these patterns can be efficiently accelerated using vector instructions.

Using these insights, we present \sysname, a technique for accelerating hashless chunking algorithms using vector instructions. 
\sysname uses a novel design to accelerate the two aforementioned patterns. 
We accelerate the extreme byte searches with a novel \textit{tree-based search} that divides a region of bytes into multiple sub-regions, processes each region using vector instructions, and uses a tree-based approach to combine their results. 
We accelerate range scans with \textit{packed scanning}, which packs multiple adjacent bytes into vector registers and compares them using a single vector operation. 

We implemented \sysname using five different vector instruction sets: SSE-128, AVX-256, and AVX-512 on Intel / AMD CPUs; NEON-128 on ARM CPUs; and VSX-128 on IBM Power CPUs. We used \sysname to accelerate three state-of-the-art hashless chunking algorithms; \newtext{\textit{RAM} \cite{ram}, \textit{AE} \cite{ae}, and \textit{MAXP} \cite{maxp}, creating \textit{VRAM}, \textit{VAE}, and \textit{VMAXP}, respectively}. We compared the performance of our accelerated algorithms with that of state-of-the-art hash-based algorithms, hashless algorithms, and SS-CDC \cite{sscdc} accelerated algorithms using 10 diverse datasets.

Our evaluation (\S\ref{sec:eval}) shows that 
\newtext{\sysname-based algorithms achieve $8.35\times$--$26.2\times$ higher chunking throughput than those accelerated with SS-CDC. \textit{VRAM}, \textit{VAE}, and \textit{VMAXP} also achieve $5.51\times$--$17.6\times$ higher throughput compared to their unaccelerated hashless counterparts, without affecting deduplication space savings. Furthermore, they achieve $15.3\times$--$207.2\times$ higher throughput compared to unaccelerated hash-based algorithms. Finally, \sysname-accelerated algorithms retain their performance advantage across all five vector instruction sets}. 

We have made our code publicly available by integrating it with DedupBench\footnote{\url{https://github.com/UWASL/dedup-bench}}~\cite{dedupbench}. Due to the large sizes of our datasets (\S\ref{sec:eval}), we were unable to release all of them. Instead, we have publicly released one of our datasets (\texttt{DEB}) on Kaggle\footnote{\url{https://www.kaggle.com/datasets/sreeharshau/vm-deb-fast25}}~\cite{deb_dataset} and provided detailed descriptions to facilitate the recreation of others for future experiments, similar to previous literature \cite{fastcdc, ae, ram}.    

The rest of this paper is organized as follows: we discuss relevant background about deduplication and vector instructions in Section \ref{sec:bg_dedup}. Section \ref{sec:motivation} motivates our work by discussing deduplication performance bottlenecks and the inefficiencies encountered by previous work when accelerating hash-based CDC algorithms. Section \ref{sec:design} outlines \sysname's design while Section \ref{sec:implementation} discusses implementation challenges across vector instruction sets. Section \ref{sec:eval} details our evaluation efforts on diverse datasets. We discuss related work in Section \ref{sec:related_work} and conclude our paper in Section \ref{sec:conclusion}.

\section{Background}
\label{sec:bg_dedup}

Data deduplication consists of four phases \cite{comprehensive_dedup}:
\begin{itemize}
    \item \textit{Data Chunking}: Data is divided into small chunks \newtext{typically of size} $1-64$KB using a chunking algorithm. All chunking algorithms have configurable parameters that control the average size of generated chunks.
    \item \textit{Fingerprinting and Comparison}: Chunks are hashed using a collision-resistant hashing algorithm such as MurmurHash3 \cite{tunable_encrypted_dedup} or SHA-256 \cite{sha256} to generate \textit{fingerprints}. Fingerprints are compared against those previously seen to identify duplicate chunks.
    \item \textit{Metadata Creation}: Metadata, i.e., \textit{file recipes} required to reconstruct the original data from stored chunks, are created.
    \item \textit{Metadata and Chunk Storage}: Non-duplicate chunks and recipes are saved on the storage medium. Fingerprints are stored on the fingerprint database and cached in an in-memory index.
\end{itemize}

Data chunking and fingerprinting are typically the most computationally intensive phases in deduplication \cite{dedup_intro}. While fingerprinting has been accelerated up to $53\times$ using GPUs \cite{gpu_dedup, storegpu} and faster hashing algorithms \cite{tunable_encrypted_dedup, persistent_memory_dedup}, data chunking acceleration has only received limited attention and adds significant overhead on the deduplication critical path
(\S\ref{sec:bg_dedup_bottlenecks}).

\subsection{Data Chunking}
\label{sec:bg_chunking} 

Data chunking algorithms can generate \textit{fixed-size} or \textit{variable-sized} chunks. Dividing the data into fixed-size chunks is fast, but results in poor space savings on most datasets (\S\ref{sec:eval}). This is due to the \textit{byte-shifting problem} \cite{lbfs}, where adding a single byte causes all subsequent chunks to appear different, despite the data stream largely being unchanged. Thus, while traditional backup systems such as Venti \cite{venti} and OceanStore \cite{oceanstore} use fixed-size chunks, modern deduplication systems employ content-defined chunking (CDC) algorithms \cite{lbfs} to generate variable-sized chunks.  

Chunk boundaries in CDC algorithms are derived from the data itself, i.e., they are \textit{content-defined}. These boundaries are chosen such that most byte shifts cause them to shift by the corresponding amount, leaving subsequent chunks unaffected and preserving the ability to detect duplication. Numerous CDC algorithms have been proposed in previous literature \cite{ae, fastcdc, gear_hash, lbfs, maxp, ram, tttd} and can be broadly classified into hash-based and hashless algorithms \cite{dedupbench}. 

\begin{figure}[t]
    \centering
     \begin{subfigure}[b]{0.45\linewidth}
        \includegraphics[width=\linewidth]{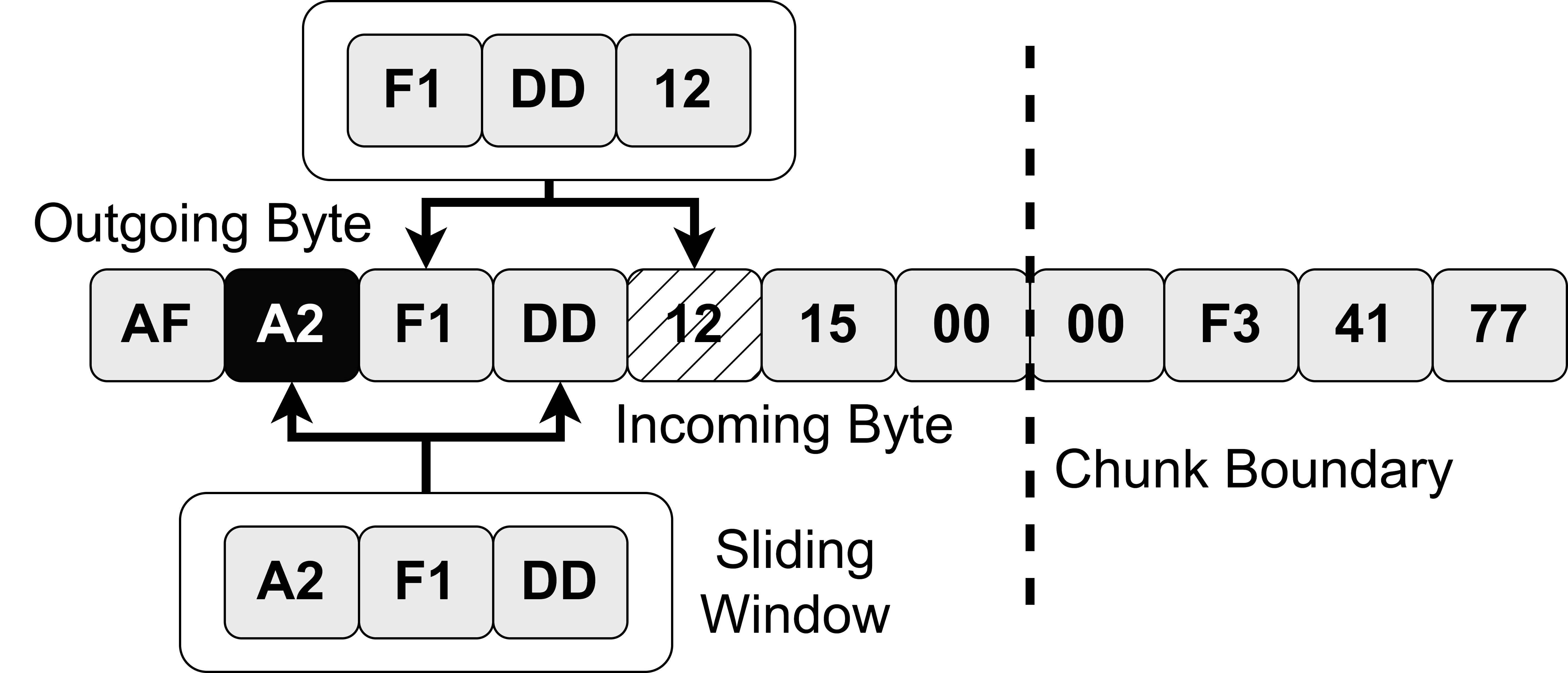}
        \caption{Sliding window and rolling hash}
        \label{fig:bg-rabin}
    \end{subfigure} \hfill
    \centering
     \begin{subfigure}[b]{0.45\linewidth}
        \includegraphics[width=\linewidth]{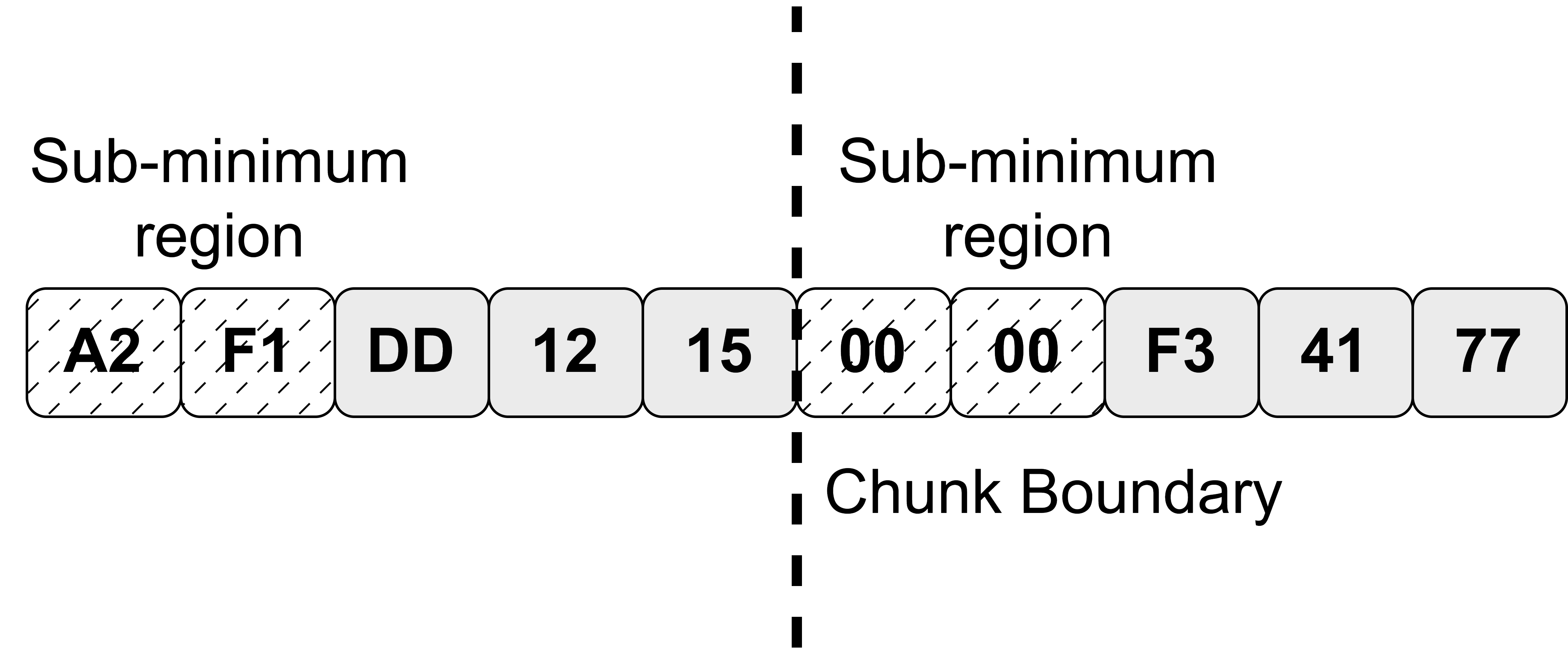}
        \caption{Minimum chunk sizes and sub-minimum skipping}
        \label{fig:bg-sub-min-hash}
    \end{subfigure}
    \caption{Hash-based chunking algorithms}
    \label{fig:bg-hash-based}
\end{figure}

\subsubsection{Hash-based algorithms.} These algorithms \cite{fastcdc, gear_hash, lbfs, tttd} slide a fixed-size window over the data. When the hash value of the window's contents matches a target mask, they insert a chunk boundary, creating a new data chunk lying between the current and previous chunk boundaries. Note that these hash-based CDC algorithms are only used during the \textit{Data Chunking} phase and do not affect the \textit{Fingerprinting and Comparison} phase.

Algorithms such as Rabin-Karp chunking \cite{lbfs} and CRC \cite{sscdc} slide a window over the source data and compute the hash of the window's contents using rolling hash algorithms. Figure \ref{fig:bg-rabin} shows an example of data chunking with such algorithms. In the Rabin-Karp chunking algorithm \cite{lbfs}, a chunk boundary is declared when the lower \textit{k} bits of the sliding window's hash value equals zero. If the current window's hash value does not meet this condition, the window is slid by a byte. To minimize the overhead of recomputing the hash value, the new value is calculated as a function of the old hash value, the incoming byte, and the outgoing byte (Figure \ref{fig:bg-rabin}), i.e., a \textit{rolling hash}. This creates a dependency between adjacent bytes, complicating acceleration efforts with SIMD instructions (\S\ref{sec:bg_sscdc}). Additionally, despite the development of more lightweight rolling hash algorithms such as CRC \cite{sscdc} and Gear-Hash \cite{gear_hash}, hash-based chunking remains computationally expensive (\S\ref{sec:eval_throughput}).

Some hash-based algorithms like TTTD \cite{tttd} and FastCDC \cite{fastcdc} use minimum and maximum values to limit the chunk sizes. To improve chunking throughput, these algorithms skip scanning data lying before the minimum chunk size at the beginning of each chunk. Figure \ref{fig:bg-sub-min-hash} shows an example of such algorithms with the sub-minimum regions highlighted using a dashed pattern. To offset the impact of skipping the sub-minimum regions and tighten chunk size distributions around the average, FastCDC \cite{fastcdc} uses dynamically changing masks, i.e., relaxes the boundary detection condition by reducing \textit{k} when required. TTTD \cite{tttd} uses two different boundary masks to do the same.

\subsubsection{Hashless algorithms.} Hashless CDC algorithms such as AE \cite{ae}, RAM \cite{ram}, and MAXP \cite{maxp} treat bytes as individual values and use local minima/maxima to identify chunk boundaries. They also slide one or more windows over the source data but do not use rolling hashes and, as a result, are faster than most hash-based algorithms by $2$--$3\times$. 

\begin{figure}[t]
    \centering
        \begin{subfigure}[b]{0.45\linewidth}
            \includegraphics[width=\linewidth]{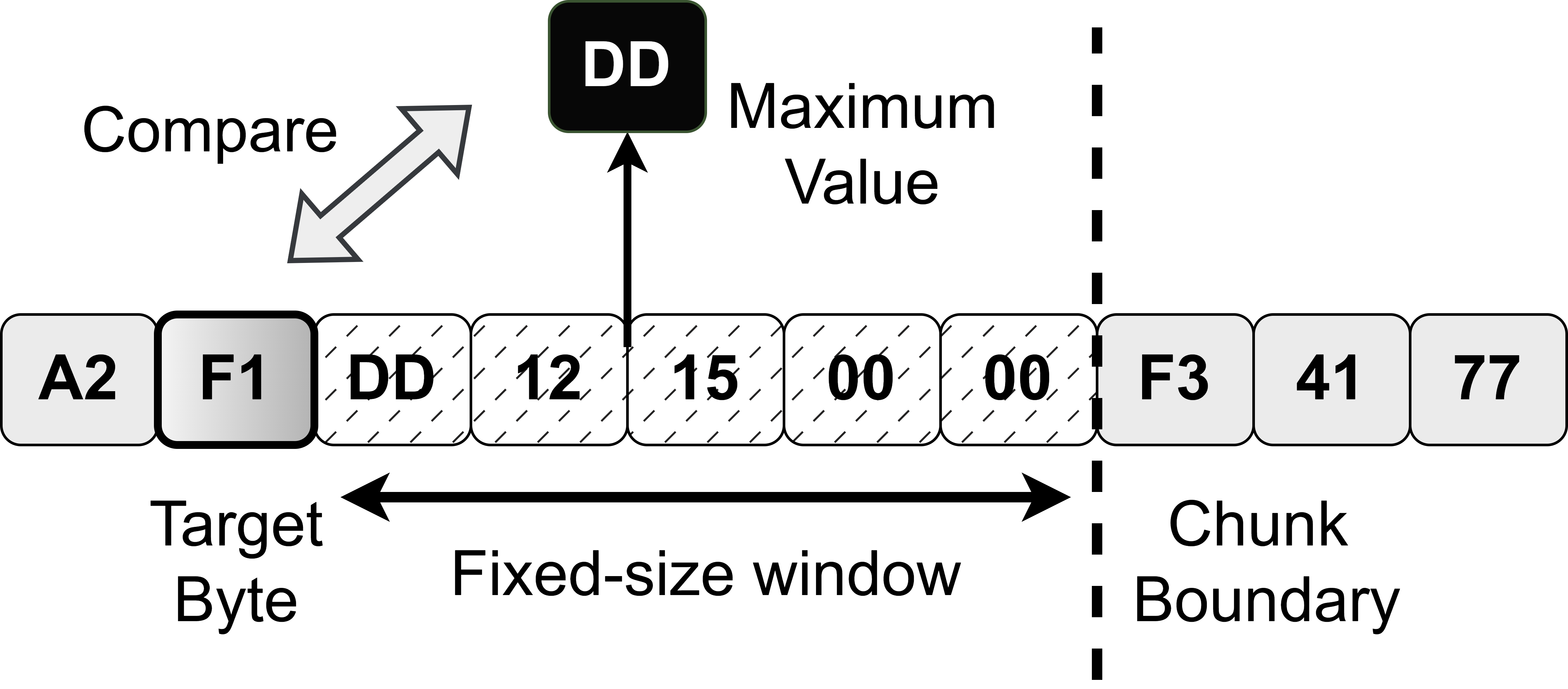}
            \caption{AE-Max}
            \label{fig:bg_ae}
        \end{subfigure}\hfill
        \begin{subfigure}[b]{0.45\linewidth}
            \includegraphics[width=\linewidth]{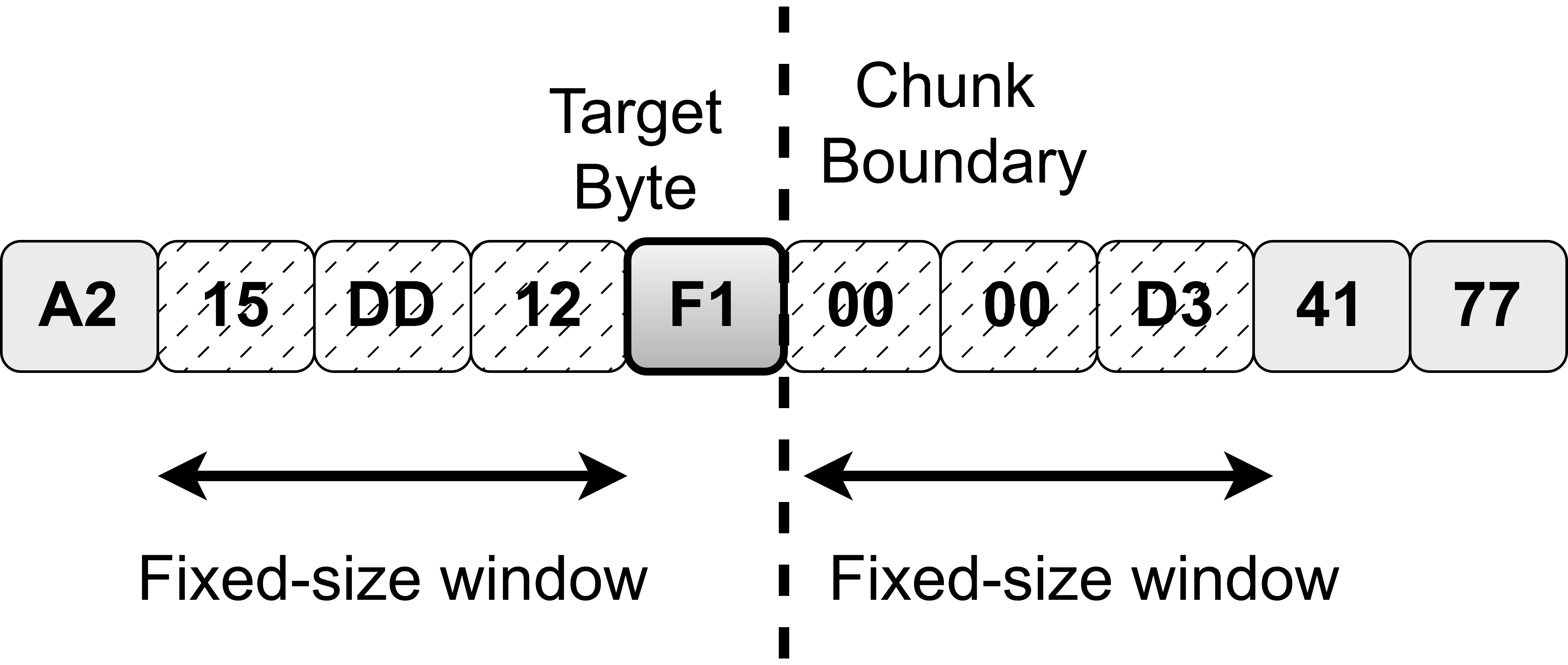}
            \caption{MAXP}
            \label{fig:bg_maxp}
       \end{subfigure}
      \begin{subfigure}[b]{0.45\linewidth}
            \includegraphics[width=\linewidth]{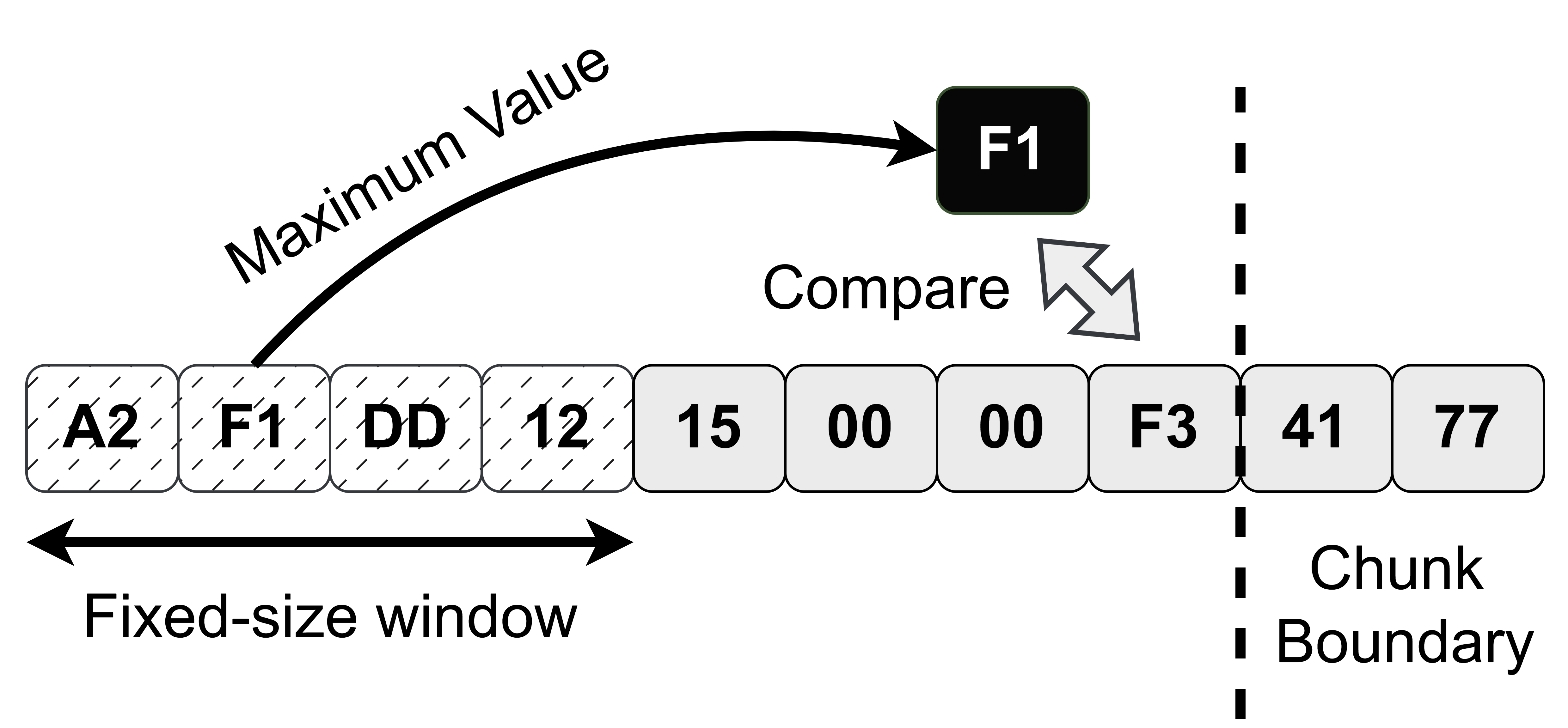}
            \caption{RAM}
            \label{fig:bg_ram}
       \end{subfigure}
    \caption{Hashless chunking algorithms}
\end{figure}

\textbf{AE.} Figure \ref{fig:bg_ae} shows an example chunk generated by the Asymmetric Extremum (AE) \cite{ae} algorithm. AE has two modes of operation: \texttt{AE-Min} and \texttt{AE-Max}, depending on whether it uses local minima or maxima; Figure \ref{fig:bg_ae} shows \texttt{AE-Max}. In each chunk, \texttt{AE-Max} tries to find a target byte that is greater than all the bytes before it. Once the target byte is identified, \texttt{AE-Max} scans a fixed-size window of bytes after the target to identify the maximum-valued byte among them. If the target byte is greater than this maximum-valued byte, it inserts a chunk boundary after the fixed-size window, as shown in Figure \ref{fig:bg_ae}.

Similarly, \texttt{AE-Min} tries to find a target byte that is less than all the bytes before it. When such a byte is identified, \texttt{AE-Min} scans a fixed-size window of bytes after the target to identify the minimum-valued byte within. If the target byte is lesser than this minimum-valued byte, it inserts a chunk boundary after the fixed-size window. 

\textbf{MAXP.} Figure \ref{fig:bg_maxp} shows an example chunk generated by MAXP \cite{maxp}. MAXP identifies target bytes in the data stream that are local maxima, i.e., they are greater than a fixed number of bytes before and after them. When such target bytes are found, chunk boundaries are inserted at their locations, as shown in the figure. Note that MAXP's window sizes are typically 70-80\% smaller than AE \cite{ae} and RAM \cite{ram} to generate the same target average chunk size. MAXP has also been referred to as Local Maximum Chunking (LMC) in previous literature.

MAXP works by sliding two fixed-size windows over the data, tracking the maximum values from both windows. These windows are located one byte apart, as shown in Figure \ref{fig:bg_maxp}, and the byte between them is the target byte. When the target byte's value is greater than the maximum value from both windows, a chunk boundary is inserted as the target byte is a local maximum. 

\textbf{RAM.} Figure \ref{fig:bg_ram} shows an example chunk generated by the Rapid Asymmetric Maximum (RAM) \cite{ram} algorithm. RAM begins by scanning a fixed-size window at the beginning of each chunk to find the maximum valued byte (\texttt{F1} in the figure). It then begins scanning at the first byte outside the window, serially comparing these bytes against this maximum value. A chunk boundary is inserted when the first byte that exceeds or equals the maximum is found, e.g., \texttt{F3} in Figure \ref{fig:bg_ram}.

As they do not possess explicit dependencies between adjacent bytes, we argue that hashless algorithms are better candidates for SIMD acceleration efforts.

\subsection{Deduplication Metrics}

Previous literature \cite{understanding_dedup_ratios, dedupbench, lbfs, fastcdc} outlines three important metrics for deduplication systems: \textit{Space Savings}, \textit{Chunk Size Distribution}, and \textit{Chunking Throughput}. We describe these in detail in this section.

\subsubsection{Space savings.} Space savings \cite{understanding_dedup_ratios, dedupbench} is one of the primary metrics used to evaluate deduplication systems in production. It represents the overall disk space conserved by using the deduplication system, i.e. size of data stored on disk after deduplication. The space savings achieved are largely dictated by the choice of the data chunking algorithm and its associated parameters. It is defined as:
\begin{equation}
    Space~Savings~(\%) = \frac{Original~Data~Size - Deduplicated~Data~Size}{Original~Data~Size} \times 100
\end{equation}


\subsubsection{Chunking throughput.} Chunking throughput is defined as the speed at which the deduplication system divides incoming data into chunks. As CDC algorithms are content-dependent, they need to scan every byte of an incoming data stream before making content-defined boundary decisions. Their speed depends on their computational complexity. Hash-based algorithms utilize expensive rolling-hash algorithms to determine chunk boundaries (\S\ref{sec:bg_chunking}), typically resulting in lower throughputs than their hashless counterparts. Chunking throughput is defined as:

\begin{equation}
    Chunking~Throughput = \frac{Original~Data~Size}{Time~taken~to~generate~all~chunk~boundaries}
\end{equation}

\subsubsection{Chunk size distribution.} Content-defined chunking (CDC) algorithms generate variable-sized chunks of a target average chunk size. They try to ensure that the sizes of generated chunks are as close to the target average as possible. However, due to underlying algorithmic characteristics, each algorithm has a unique chunk size distribution pattern. For instance, FastCDC \cite{fastcdc} exhibits two distinct smooth distributions, changing its pattern at the target average chunk size. This is because it switches masks past the target average size and relaxes boundary conditions. On the other hand, algorithms such as TTTD \cite{tttd} exhibit a smooth distribution between their minimum and maximum specified chunk sizes. Chunk size distributions are typically represented using cumulative distribution function (CDF) plots. 

Space savings are inversely proportional to the target average chunk size, i.e., the greater the average chunk size, the lower the space savings achieved in general \cite{primary_dedup}. This is because the probability of finding duplicate chunks is higher at smaller chunk sizes. The degree of space savings degradation with increasing chunk size depends on algorithmic and dataset characteristics. 


All chunks generated by CDC algorithms are subsequently hashed using a collision-resistant algorithm \cite{tunable_encrypted_dedup} to generate fingerprints, as described above. The set of unique fingerprints observed thus far is stored in a \textit{fingerprint database}. New incoming chunks are hashed, and their fingerprints are compared against this database to detect duplicates. Thus, smaller and more numerous chunks result in a larger database; specifically, the fingerprint database size is inversely proportional to the chosen average chunk size.  A large number of small chunks can negatively impact system throughput, both due to the increased fingerprint database size and the random data accesses caused by these chunks. Thus, CDC algorithms in production typically target average chunk sizes between $2$KB--$64$KB.

\subsection{Vector Instructions}
\label{sec:bg_vector_inst}

Vector instructions \cite{vector_instruction_sets} are special Single-Instruction Multiple-Data (SIMD) instructions supported by most modern processors. They rely on special vector registers for their operations. These registers come in multiple sizes; depending on the width of the vector register they use, vector instructions can be classified into different families \cite{vector_instruction_sets}. The most common vector register sizes are 128 bits, 256 bits, and 512 bits, i.e., 16, 32, and 64 bytes wide. 

Vector instructions support the execution of an operation on multiple pieces of data by packing them into vector registers; for instance, eight 16-bit values $a$-$h$ can be densely packed into a 128-bit vector register $V_{1}$. To add $a$-$h$ to eight other values $i$-$p$, we can pack $i$-$p$ into another register $V_{2}$. We can now add them pairwise with a single vector addition operation \textit{VADD~($V_{1}, V_{2}$)} using $V_{1}$ and $V_{2}$ as operands, instead of eight separate integer arithmetic operations.

Vector instructions support various arithmetic operations \cite{intelIntrinsicsGuide}, including pairwise addition, subtraction, multiplication, and maximum/minimum on packed values. Additionally, they support logical operations such as bitwise AND (\texttt{\&}) and bitwise OR (\texttt{|}). They have been previously used to accelerate matrix multiplication \cite{avx_matrixmul}, sorting \cite{avx_quicksort}, multimedia applications \cite{avx_multimedia}, fluid simulations \cite{avx_fluidmech}, hash tables \cite{hashtables_vector}, and relational databases \cite{fused_table_scan}. The supported vector instruction types and their relative performance vary across CPU manufacturers.

\subsubsection{Intel and AMD} Vector instructions on x86 platforms can be classified into three families: SSE-128, AVX-256, and AVX-512 \cite{vector_instruction_sets}. SSE-128 instructions use 128-bit registers and have been supported by Intel and AMD processors since 1999 \cite{intelSIMDInstructions} and 2003 \cite{AMD_SSE128}, respectively. AVX-256 instructions were introduced by Intel and AMD in 2011 \cite{intelSIMDInstructions}, and use 256-bit registers. Finally, only a handful of the newest Intel and AMD processors, since 2017 \cite{skylake_wikichip} and 2022 \cite{amdzen4_wikichip}, which have 512-bit wide vector registers support AVX-512 instructions.  

\subsubsection{ARM} ARM processors have supported NEON-128 instructions, an equivalent to SSE-128, since 2011 \cite{armNeonOverview}. Modern ARM processors also support vector widths of 256 bits and higher with the SVE/SVE2 instruction sets, which have been available since 2021 \cite{arm_sve}. These two instruction sets differ in the kinds of instructions supported. For instance, NEON-128 does not support native \textit{VMASK} operations, which are used to create integer masks by extracting one out of every \textit{k} bits in a vector register, while SVE / SVE2 does. This can lead to performance differences in applications that need the \textit{VMASK} operation \cite{hashtables_vector}.

\subsubsection{IBM Power} IBM's Power \cite{ibm_power8} architecture supports AltiVec / VSX-128 vector instructions \cite{altivec_instructions}, an equivalent to SSE-128, since the 1990s. This instruction set supports equivalents for most SSE instructions but lacks \textit{VMASK} support.

\section{Motivation}
\label{sec:motivation}

\subsection{Performance bottlenecks in data deduplication}
\label{sec:bg_dedup_bottlenecks}

\begin{figure}[t]
    \begin{subfigure}[b]{0.8\linewidth}
    \centering
        \includesvg[inkscapelatex=false,width=\linewidth]{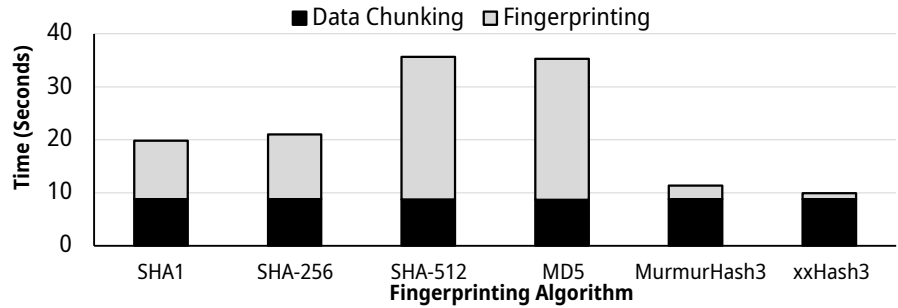}
    \end{subfigure}
    \caption{Time taken for Data Chunking vs Fingerprinting while deduplicating randomized data with FastCDC \cite{fastcdc} and an 8 KB average chunk size.}
    \label{fig:chunking_hashing_time_motivation}
\end{figure}

Although both data chunking and fingerprinting have traditionally been considered the main bottlenecks in deduplication \cite{dedup_intro}, this has changed with the advent of new hashing algorithms and acceleration methods for fingerprinting. Figure \ref{fig:chunking_hashing_time_motivation} is a stacked bar plot showing the time taken by the data chunking and fingerprinting phases during deduplication. For fingerprinting, we use five different collision-resistant hashing algorithms \cite{md5, sha1, sha256_512, appleby2008murmurhash, hashing_survey_xxhash}. For data chunking, we use FastCDC \cite{fastcdc}, the fastest unaccelerated chunking algorithm (\S\ref{sec:eval_throughput}), with an 8 KB average chunk size. We use 30 GB of randomized data and an Intel Emerald Rapids machine described in \S\ref{sec:eval} for this experiment.  

Figure \ref{fig:chunking_hashing_time_motivation} shows that fingerprinting and data chunking take nearly equal time with traditional collision-resistant hashing algorithms, such as SHA1 \cite{sha1} and SHA-256 \cite{sha256_512}. Fingerprinting takes longer than data chunking with other algorithms, such as MD5 \cite{md5} and SHA-512 \cite{sha256_512}. This indicates that both phases used to be the performance bottlenecks in deduplication, aligning with previous literature \cite{dedup_intro}.

However, recent research has introduced faster hashing algorithms such as MurmurHash3 \cite{appleby2008murmurhash} and xxHash3 \cite{xxhashWebsite} \newtext{that generate 128-bit digests equivalent to MD5 \cite{appleby2016smhasher}}. \newtext{These hashing algorithms are being used for fingerprinting \cite{scalable_incremental_checkpt} or as weak fingerprints followed by a byte-by-byte comparison \cite{xxhash_weak_ssd}}. Figure \ref{fig:chunking_hashing_time_motivation} shows that fingerprinting takes significantly lower time than chunking with these new hashing algorithms, as they are $10\times-15\times$ faster than their counterparts on CPUs. Using GPUs can further accelerate fingerprinting speeds by up to $53\times$ \cite{gpu_dedup, storegpu}. 

Thus, as a result of its computationally intensive nature and position on the critical path, \textit{data chunking is \newtext{a prime target for acceleration.}}

\subsection{Accelerating hash-based algorithms with vector instructions}
\label{sec:bg_sscdc}

\begin{figure}[t]
    \begin{subfigure}[b]{0.6\linewidth}
    \centering
        \includegraphics[width=\linewidth]{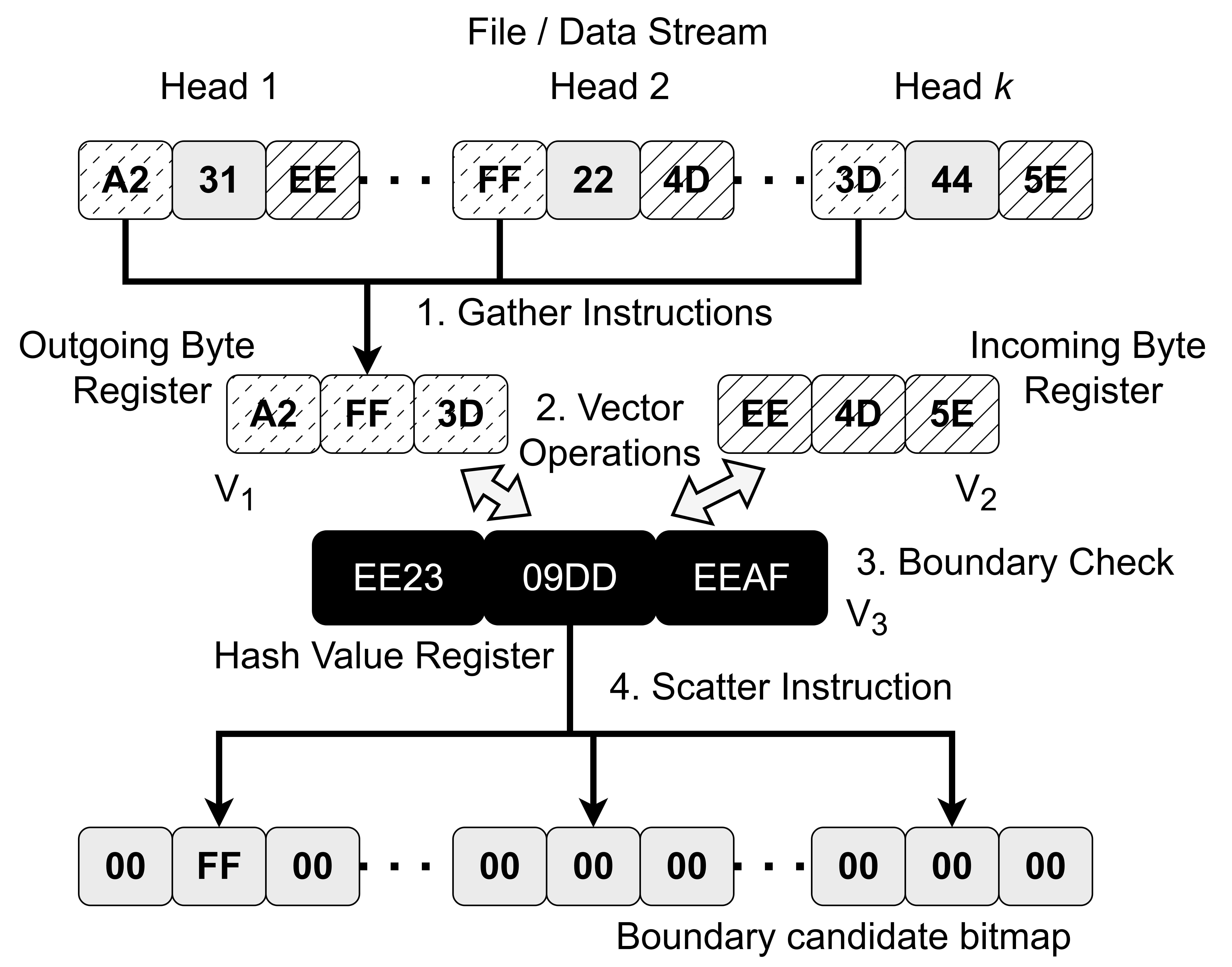}
    \end{subfigure}
    \caption{\newtext{SS-CDC \cite{sscdc}: Accelerating the rolling hash phase}}
    \label{fig:ss_cdc_motivation}
\end{figure}

To address the data chunking bottleneck, SS-CDC \cite{sscdc} proposed using AVX-512 instructions to accelerate hash-based CDC algorithms. They decouple the rolling hash and boundary detection phases, running the rolling hash on the entire source data to identify boundary candidates in the first phase, and determining boundaries sequentially in the second. This allows both stages to be independently accelerated with AVX instructions. 

Figure \ref{fig:ss_cdc_motivation} shows how SS-CDC \cite{sscdc} accelerates the first rolling hash phase of hash-based CDC algorithms. SS-CDC uses AVX-512 registers to create multiple \textit{rolling-heads} \newtext{(Head 1 - Head \textit{k})}, i.e., calculating the rolling hash on bytes from multiple regions of the file independently and in parallel. Each rolling head maintains its own hash value \newtext{in the hash value register} and independently calculates the contributions of incoming and outgoing bytes. 

To use vector instructions, they first collect the outgoing bytes for each head into a vector register $V_1$. Similarly, they collect all the incoming bytes into another vector register $V_2$. \newtext{This is shown in Step 1 in Figure \ref{fig:ss_cdc_motivation} and uses \texttt{gather} instructions.} The hash values for each head are stored in a separate register $V_3$. \newtext{In Step 2,} SS-CDC removes the contributions of all outgoing bytes from the hash values with a single vector operation and adds the contributions of all incoming bytes with another. 

\newtext{Step 3 compares all the hash values against the pre-specified boundary condition (such as the lower \textit{x} bits being equal to zero in Rabin-Karp chunking \cite{lbfs})}. Whenever any of the hash values match the boundary condition, they mark the current position as a boundary candidate in a separate bitmap \newtext{in Step 4 using \texttt{scatter} instructions. This rolling hash phase is run on the entire incoming data stream/file.} In the second phase, they scan the bitmap using vector instructions to determine the actual boundaries among all candidates, taking into account the minimum and maximum chunk sizes.

This approach introduces two problems. First, many hash-based algorithms, such as TTTD and FastCDC, skip scanning data up to the minimum chunk size to improve throughput (\S\ref{sec:bg_chunking}). Decoupling the rolling hash and boundary detection phases causes the rolling hash to be run on the entire incoming data stream, nullifying these optimizations. 

Second, to load incoming and outgoing bytes from different regions in the file, SS-CDC \cite{sscdc} uses AVX \texttt{gather} instructions. To populate the candidate bitmap when boundary candidates are discovered, they use \texttt{scatter} instructions. These \texttt{scatter} and \texttt{gather} instructions are slow \cite{scatter_gather_perf}, limiting performance gains. Finally, \texttt{scatter} instructions are only available on processors supporting certain instruction sets \cite{intelIntrinsicsGuide}, limiting SS-CDC's usage to a handful of the newest Intel and AMD processors (\S\ref{sec:bg_vector_inst}).

\begin{figure}[t]
    \centering
     \begin{subfigure}[b]{0.99\linewidth}
     \centering
        \includegraphics[width=0.45\linewidth]{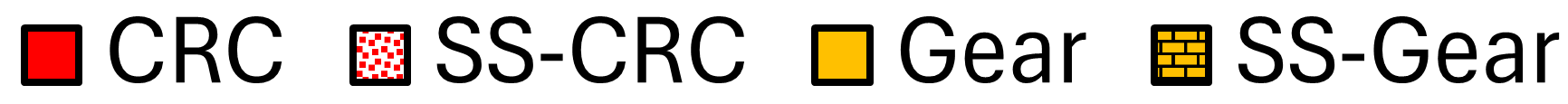}
    \end{subfigure}
    \begin{subfigure}[b]{0.6\linewidth}
        \includesvg[inkscapelatex=false,width=\linewidth]{figures/Motivation_Throughput.svg}
    \end{subfigure}
    \caption{SS-CDC \cite{sscdc} throughputs on randomized data with AVX-512 instructions}
    \label{fig:throughput_motivation}
\end{figure}

Figure \ref{fig:throughput_motivation} shows the chunking throughput obtained by running SS-CDC accelerated versions of CRC (\textit{SS-CRC}) and Gear-based chunking (\textit{SS-Gear}) \cite{sscdc} against their native unaccelerated counterparts. This experiment used randomized data, an Intel Emerald Rapids machine described in \S\ref{sec:eval} and AVX-512 instructions. We ran each algorithm with chunk sizes of $4-16$ KB. \textit{SS-CRC} achieves 1.2 GB/s, a speedup of $2.58\times$ over \textit{CRC}. Similarly, \textit{SS-Gear} achieves 3.3 GB/s, a speedup of $3.13\times$ over \textit{Gear}. These small speedups result from the challenges associated with hash-based algorithms that are described above. 

Hashless algorithms do not possess explicit dependencies between adjacent bytes. They treat each byte as an independent value and use maximum / minimum values from data regions to determine chunk boundaries. \sysname chooses hashless algorithms over their hash-based counterparts as they are better candidates for SIMD acceleration.


\section{\sysname Design}
\label{sec:design}

Hashless CDC algorithms such as \textit{AE} \cite{ae}, \textit{RAM} \cite{ram}, and \textit{MAXP} \cite{maxp} slide windows over the source data to determine chunk boundaries. We identify two common processing patterns across all hashless CDC algorithms: the \textit{Extreme Byte Search} and \textit{Range Scan}. We accelerate each of these patterns using different vector-based techniques, which are discussed in detail below.

While we use the AVX-512 instruction set as an example to describe our acceleration techniques in this section, they can be implemented on any CPU with a vector instruction set supporting \textit{VMAX}, \textit{VCMP} and \textit{VMASK} operations. SSE-128 and AVX-256 instruction sets \cite{intelIntrinsicsGuide} fall under this umbrella, as do ARM and IBM processors with NEON-128 \cite{armNeonOverview} and AltiVec / VSX-128 \cite{ibm_power8} instructions, respectively. Thus, \sysname is compatible with a wide range of processors, unlike SS-CDC \cite{sscdc}, which relies on \texttt{scatter} instructions only available in AVX-512 instruction sets. Finally, while other minima/maxima-based hashless algorithms can also be accelerated using \sysname, their native versions are slower \cite{ae, ram, maxp} than \textit{AE}, \textit{RAM}, and \textit{MAXP} and have been omitted from the rest of our paper. 


\begin{figure}[t]
    \centering
    \includegraphics[width=0.67\linewidth]{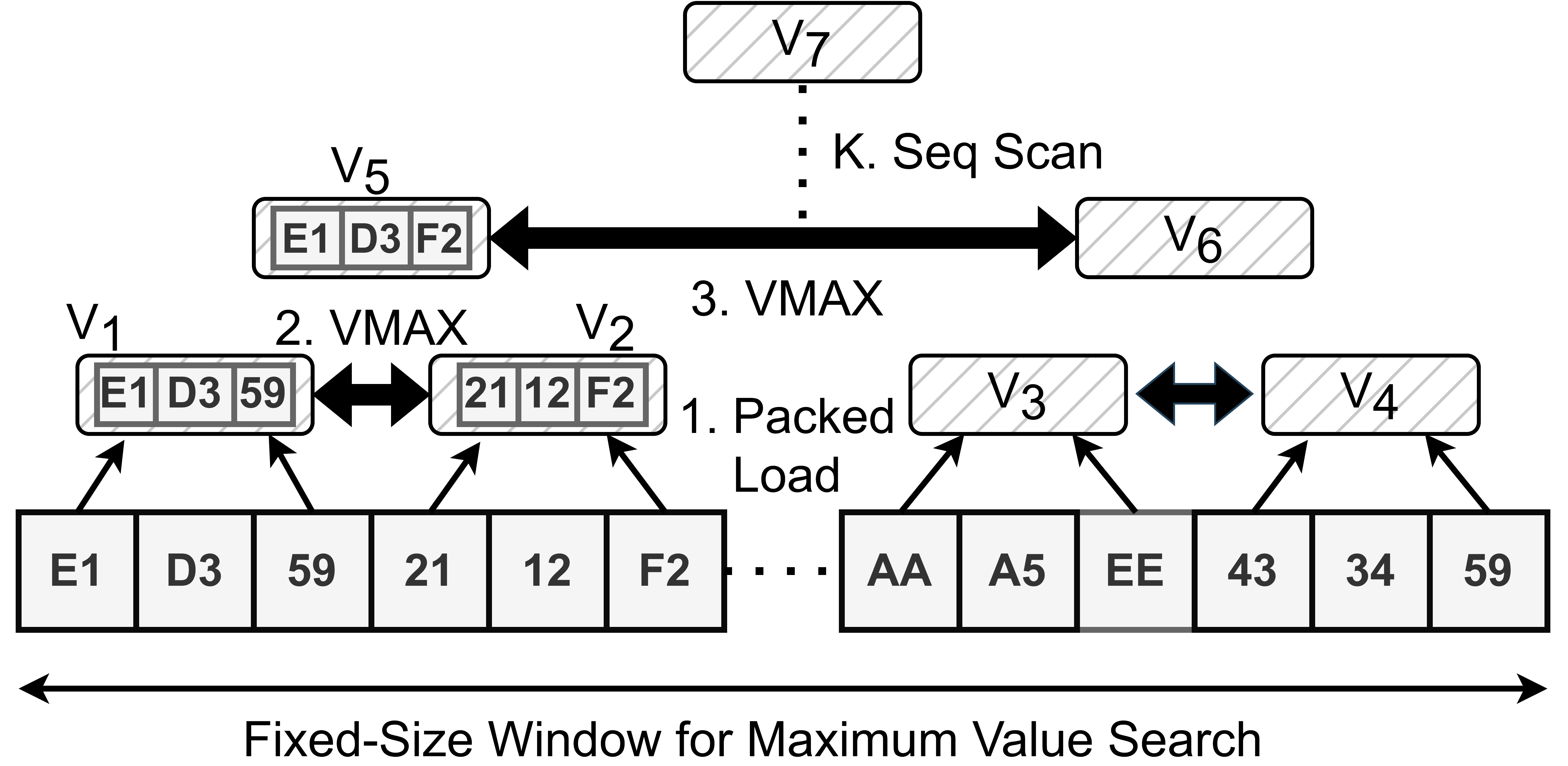}
    \caption{Accelerating Extreme Byte Searches. Note that the byte values shown are in hexadecimal format.}
    \label{fig:design_phase1}
\end{figure}

\subsection{Tree-based Extreme Byte Search}

Hashless CDC algorithms such as \textit{AE} \cite{ae}, \textit{RAM} \cite{ram}, and \textit{MAXP} \cite{maxp} all consist of a subsequence that identifies the \textit{extreme byte} (maximum/minimum) in a fixed-size window. The size of this window depends upon the expected average chunk size and can be as large as $4-8$KB. As this subsequence may need to be performed more than once per chunk, we propose accelerating it using a novel \textit{tree-based search approach}. Let us consider the search for a maximum value using AVX-512 instructions (Figure \ref{fig:design_phase1}). Note that the same method can be used with other vector instruction sets as well as to find minimum values.

We first divide the fixed-size window into smaller sub-regions, loading all the bytes into AVX-compatible \texttt{m512i} variables in \textit{Step 1}. We load these bytes in a packed fashion i.e. each \texttt{m512i} variable contains 64 adjacent bytes. We then use vector \texttt{mm512\_max} instructions to find the pairwise maximum among packed byte pairs (\textit{Step 2}). For instance, among the bytes \texttt{0xE1} and \texttt{0x21}, byte value \texttt{0xE1} is the maximum. The resulting pairwise maximums are packed into a destination variable ($V_5$ in the figure). 

\textit{Step 3} compares these resulting variables $V_5$ and $V_6$ from \textit{Step 2} using \texttt{mm512\_max} instructions to find the pairwise maximums. We repeat this process, building a tree of \texttt{m512i} variables until we are left with a single variable $V_7$ containing the maximum-valued 64 bytes from across the entire region. We scan these bytes sequentially in \textit{Step K} to determine the maximum valued byte. 


\subsection{Packed Scanning for Range Scans}

\begin{figure}[t]
    \centering
    \includegraphics[width=0.67\linewidth]{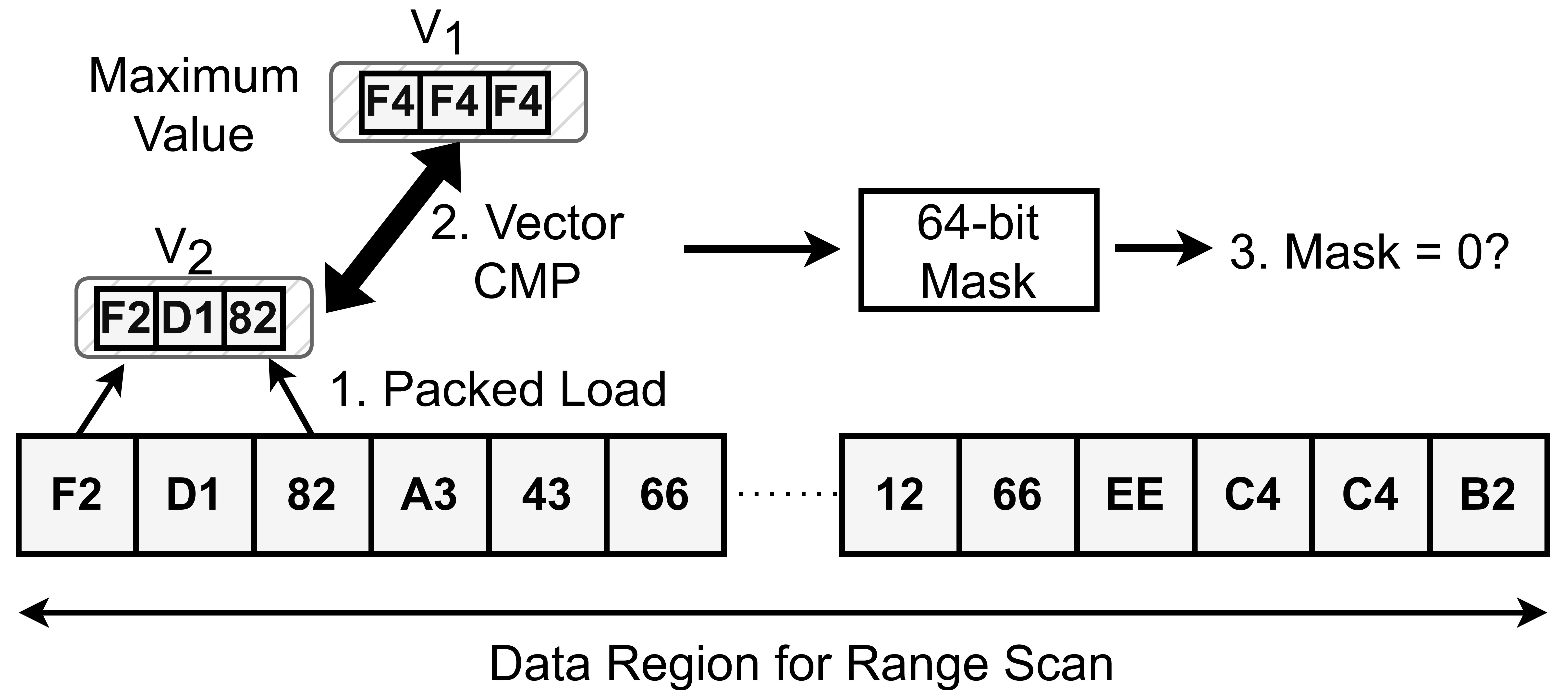}
    \caption{Accelerating Range Scans. Note that the byte values shown are in hexadecimal format.}
    \label{fig:design_phase2}
\end{figure}

Hashless CDC algorithms also consist of a range scan subsequence, where bytes are serially compared against a target value. We propose to accelerate this scanning process using vector instructions. Let us consider a case where we compare bytes sequentially to see if they are greater than or equal to a target value (such as in \textit{RAM} \cite{ram}). Figure \ref{fig:design_phase2} shows our proposal to accelerate this using \textit{packed scanning} with AVX-512 instructions. Note that the same methods are applicable for other vector instruction sets as well.

We first load the target value (\texttt{0xF4} in Figure \ref{fig:design_phase2}) into an AVX-compatible \texttt{m512i} variable $V_1$. We then pack 64 adjacent bytes from the scan region into another \texttt{m512i} variable $V_2$. We compare these 2 registers using \texttt{mm512\_cmpge} vector compare instructions, which generate a 64-bit integer mask containing the comparison results. If this mask has a value greater than 0, a chunk boundary exists within the scanned 64 bytes. Its exact position is determined using the mask value. If the mask equals 0, no boundary exists within the scanned region and we proceed with loading the next 64 bytes into $V_2$ to repeat the process. 

\textit{Range Scans} can be run with one of five comparators: Greater-Than (\textit{GT}), Lesser-Than (\textit{LT}), Greater-Than or Equals (\textit{GEQ}), Less-Than or Equals (\textit{LEQ}), and exactly Equals (\textit{EQ}). Each of these comparators uses a different vector compare instruction; for instance, the \textit{GEQ} comparator uses \texttt{mm512\_cmpge} instructions while the \textit{LEQ} comparator uses \texttt{mm512\_cmple} instructions. The same comparators also use different comparison instructions in different instruction sets; for instance, the \textit{GEQ} comparator uses \texttt{mm512\_cmpge} instructions with the AVX-512 instruction set while it uses \texttt{mm256\_cmpge} with AVX-256.

It is worth noting that our \textit{packed scanning} approach is compatible with sub-minimum skipping. Unlike SS-CDC's approach, chunk boundary detection and insertion can both occur in \textit{Range Scans}, i.e., whenever a chunk boundary is detected, the next \texttt{minimum\_chunk\_size} bytes can be skipped.

\subsection {Putting it together: AE-Max, AE-Min, MAXP, and RAM}
\label {sec:design_combine}

\textit{RAM} \cite{ram} first scans a fixed-size window at the beginning of the chunk to find a maximum value (Figure \ref{fig:bg_ram}). After this, it inserts a chunk boundary at the first byte outside the window, which is at least as large as the maximum valued byte (\S\ref{sec:bg_chunking}). With \sysname, we accelerate \textit{RAM} as a combination of an \textit{Extreme Byte Search} to find a maximum value, followed by a \textit{Range Scan} with the \textit{GEQ} comparator that compares this maximum value against bytes until a chunk boundary is found.

\textit{AE-Max} \cite{ae} scans for a byte larger than all the bytes before it i.e., a target byte (Figure \ref{fig:bg_ae}). Once found, a fixed-size window after this byte is scanned to determine the maximum valued byte within. If the target byte is larger than the maximum valued byte, a chunk boundary is inserted; otherwise, scanning continues for a new target byte (\S\ref{sec:bg_chunking}). With \sysname, we accelerate \textit{AE-Max} as a combination of multiple \textit{Range Scans} with the \textit{GT} comparator to find target bytes, each followed by a single \textit{Extreme Byte Search} for a maximum value. 

\textit{AE-Min} \cite{ae} scans for a byte with lesser value than all those before it (\S\ref{sec:bg_chunking}). Once found, a fixed-size window after this byte is scanned to determine the minimum value within. If the target byte has a lesser value than the minimum valued byte, a chunk boundary is inserted; otherwise, scanning continues for a new target byte. Similar to \textit{AE-Max}, we accelerate \textit{AE-Min} as a combination of multiple \textit{Range Scans} with the \textit{LT} comparator to find target bytes, each followed by a single \textit{Extreme Byte Search} for a minimum value.

Finally, \textit{MAXP} \cite{maxp} scans for a target local maxima that is exactly centered between two fixed-size windows (Figure \ref{fig:bg_maxp}). A chunk boundary is inserted right after such a byte is found (\S\ref{sec:bg_chunking}). Thus, each chunk in \textit{MAXP} can be represented as a combination of multiple \textit{Range Scans} with the \textit{GT} comparator, each followed by two \textit{Extreme Byte Searches} for maximum values.


\section{Implementation}
\label{sec:implementation}


We accelerate \textit{AE} \cite{ae}, \textit{MAXP} \cite{maxp}, and \textit{RAM} \cite{ram} using \sysname with 3000 lines of C++ code. We implemented SSE-128, AVX-256, AVX-512, NEON-128, and VSX-128 versions of all algorithms. We also implemented \textit{Extreme Byte Searches} for minima and maxima, as well as \textit{Range Scan} functionalities with the \textit{GT}, \textit{GEQ}, \textit{LT}, \textit{LEQ}, and \textit{EQ} comparators on all five vector instruction sets. We have made our code publicly available with DedupBench\footnote{\url{https://github.com/UWASL/dedup-bench}}~\cite{dedupbench}.

Note that while ARM processors support \textit{VCMP} and \textit{VMAX} operations, they lack native support for \textit{VMASK} instructions, which are used during range scans to generate a single mask containing the comparison results. This is a common issue encountered by ARM developers trying to port x86 code \cite{armPortingVector}. We chose an efficient alternative implementation \cite{armPortingVector} to work around the lack of native \textit{VMASK} support. However, this alternative implementation uses multiple slow NEON-128 instructions, such as \texttt{vshrn} and \texttt{vreinterpretq}, as opposed to a single x86 \texttt{mm\_movemask} instruction. As shown in \S\ref{sec:eval_proc_compat}, this causes accelerated algorithms to achieve lower speedups on ARM CPUs compared to Intel and AMD.

IBM processors also support \textit{VCMP} and \textit{VMAX} operations, but lack native \textit{VMASK} support. However, the same functionality can be achieved using one \texttt{vec\_bperm} and two \texttt{vec\_extract} instructions. As these instructions are relatively inexpensive, they are an efficient alternative to \textit{VMASK}. As shown in \S\ref{sec:eval_proc_compat}, this allows IBM processors to achieve speedups equivalent to or greater than Intel and AMD processors when using \sysname.
\section{Evaluation}
\label{sec:eval}

In this section, we evaluate \sysname against the state-of-the-art CDC algorithms. 

\begin{table}[t]
\centering
\renewcommand{\arraystretch}{1.25}
\begin{tabular}{cccccc}
\rowcolor[HTML]{000000} 
\multicolumn{1}{l}{\cellcolor[HTML]{000000}{\color[HTML]{FFFFFF} \textbf{CPU / CPU Family}}} &
  \multicolumn{1}{l}{\cellcolor[HTML]{000000}{\color[HTML]{FFFFFF} \textbf{SSE-128}}} &
  \multicolumn{1}{l}{\cellcolor[HTML]{000000}{\color[HTML]{FFFFFF} \textbf{AVX-256}}} &
  \multicolumn{1}{l}{\cellcolor[HTML]{000000}{\color[HTML]{FFFFFF} \textbf{AVX-512}}} &
  \multicolumn{1}{l}{\cellcolor[HTML]{000000}{\color[HTML]{FFFFFF} \textbf{NEON-128}}} &
  \multicolumn{1}{l}{\cellcolor[HTML]{000000}{\color[HTML]{FFFFFF} \textbf{VSX-128}}} \\ \hline
\multicolumn{1}{|c|}{Intel Emerald Rapids} &
  \multicolumn{1}{c|}{{\color[HTML]{000000} \cmark}} &
  \multicolumn{1}{c|}{{\color[HTML]{000000} \cmark}} &
  \multicolumn{1}{c|}{{\color[HTML]{000000} \cmark}} &
  \multicolumn{1}{c|}{{\color[HTML]{000000} \textbf{--}}} &
  \multicolumn{1}{c|}{{\color[HTML]{000000} \textbf{--}}} \\ \hline
\multicolumn{1}{|c|}{Intel Skylake} &
  \multicolumn{1}{c|}{{\color[HTML]{000000} \cmark}} &
  \multicolumn{1}{c|}{{\color[HTML]{000000} \cmark}} &
  \multicolumn{1}{c|}{{\color[HTML]{000000} \cmark}} &
  \multicolumn{1}{c|}{{\color[HTML]{000000} \textbf{--}}} &
  \multicolumn{1}{c|}{{\color[HTML]{000000} \textbf{--}}} \\ \hline
\multicolumn{1}{|c|}{AMD EPYC Rome} &
  \multicolumn{1}{c|}{{\color[HTML]{000000} \cmark}} &
  \multicolumn{1}{c|}{{\color[HTML]{000000} \cmark}} &
  \multicolumn{1}{c|}{{\color[HTML]{000000} \textbf{--}}} &
  \multicolumn{1}{c|}{{\color[HTML]{000000} \textbf{--}}} &
  \multicolumn{1}{c|}{{\color[HTML]{000000} \textbf{--}}} \\ \hline
\multicolumn{1}{|c|}{ARM v8 Atlas} &
  \multicolumn{1}{c|}{{\color[HTML]{000000} \textbf{--}}} &
  \multicolumn{1}{c|}{{\color[HTML]{000000} \textbf{--}}} &
  \multicolumn{1}{c|}{{\color[HTML]{000000} \textbf{--}}} &
  \multicolumn{1}{c|}{{\color[HTML]{000000} \cmark}} &
  \multicolumn{1}{c|}{{\color[HTML]{000000} \textbf{--}}} \\ \hline
\multicolumn{1}{|c|}{IBM Power 8} &
  \multicolumn{1}{c|}{{\color[HTML]{000000} \textbf{--}}} &
  \multicolumn{1}{c|}{{\color[HTML]{000000} \textbf{--}}} &
  \multicolumn{1}{c|}{{\color[HTML]{000000} \textbf{--}}} &
  \multicolumn{1}{c|}{{\color[HTML]{000000} \textbf{--}}} &
  \multicolumn{1}{c|}{{\color[HTML]{000000} \cmark}} \\ \hline
\end{tabular}
\caption{Vector instruction sets supported by the different machines in our testbed.}
\label{tbl:cpu_inst_support}
\end{table}

\textbf{Testbed.} We run all our experiments using machines from the Cloudlab \cite{cloudlab_paper} platform. We pick five machines with diverse vector instruction set support; Table \ref{tbl:cpu_inst_support} shows the vector instruction sets supported by each machine.  The details of each machine are as follows:
\begin{itemize}
    \setlength\itemsep{0.3em}
       \item \textit{Intel Emerald Rapids:} We use a \textit{c6620} machine from CloudLab Utah, which has a 28-core Intel Xeon Gold 5512U with hyperthreading at 2.1 GHz, 128 GB of RAM, and one Intel NIC each of 25 GBps and 100 GBps. It supports the SSE-128, AVX-256, and AVX-512 vector instruction sets.
    \item \textit{Intel Skylake:} We use a \textit{c240g5} machine from CloudLab Wisconsin, which has two 10-core Intel Xeon Silver 4114 CPUs with hyperthreading at 2.2 GHz, 192 GB of RAM, one Mellanox 25 GBps NIC, and one onboard Intel 1 GBps NIC. It supports the SSE-128, AVX-256, and AVX-512 vector instruction sets.
    \item \textit{AMD EPYC Rome:} We use a \textit{c6525-25g} machine from CloudLab Utah, which has a 16-core AMD 7302P CPU with hyperthreading at 3.0 GHz, 128 GB of RAM, and two Mellanox 25 GBps NICs. It supports the SSE-128 and AVX-256 vector instruction sets.
    \item \textit{ARM v8 Atlas:} We use a \textit{m400} machine from CloudLab Utah, which has an 8-core ARM Cortex A-57 CPU at 2.4 GHz, 64 GB of RAM, and a 10 GBps Mellanox NIC. It supports the NEON-128 vector instruction set.
    \item \textit{IBM Power 8:} We use an \textit{ibm8335} machine from CloudLab Clemson, which has dual 10-core IBM Power8NVL CPUs at 2.86 GHz with 8 hardware threads per core, 256 GB of RAM, and a 10 GBps Broadcom Xtreme II NIC. It supports the VSX-128 vector instruction set.  
\end{itemize}


While some ARM CPUs released after 2022 support higher vector widths with SVE instructions (\S\ref{sec:bg_vector_inst}), we could not obtain such a machine for our experiments. Note that all our runs are on the Intel Emerald Rapids machine unless otherwise specified. Our throughput results are the averages of 5 runs, and the standard deviation was less than $5\%$.

\textbf{Alternatives.} We evaluate the following hash-based CDC algorithms:
\begin{itemize}
    \setlength\itemsep{0.3em}
    \item {\textit{CRC:}} Native (unaccelerated) version of the CRC-32 chunking algorithm from SS-CDC \cite{sscdc}.
    \item {\textit{FCDC:}} Native version of FastCDC \cite{fastcdc}.
    \item {\textit{Gear:}} Native version of the Gear-hash based chunking algorithm \cite{gear_hash}.
    \item {\textit{RC:}} Rabin's chunking algorithm from LBFS \cite{lbfs}.
    \item {\textit{SS-CRC:}} AVX-512 version of CRC accelerated using SS-CDC \cite{sscdc}. 
    \item {\textit{SS-Gear:}} AVX-512 version of Gear accelerated using SS-CDC \cite{sscdc}. 
    \item {\textit{TTTD:}} Two-Threshold Two-Divisor algorithm \cite{tttd}.
\end{itemize}

We also evaluate the following hashless CDC algorithms:
\begin{itemize}
    \setlength\itemsep{0.3em}
    \item {\textit{AE:}} Native version of the Asymmetric Extremum algorithm \cite{ae}. We evaluate both \textit{AE-Max} and \textit{AE-Min}.
    \item {\textit{MAXP:}} Native version of the MAXP algorithm \cite{maxp}.
    \item {\textit{RAM:}} The native Rapid Asymmetric Maximum \cite{ram} algorithm.
    \item {\textit{VAE}:} Accelerated versions of \textit{AE-Max} and \textit{AE-Min} with \sysname.
    \item {\textit{VMAXP}:} Accelerated versions of MAXP with \sysname.
    \item {\textit{VRAM}:} Accelerated versions of RAM with \sysname.
\end{itemize}

Note that for each hashless algorithm accelerated with \sysname, we evaluate their SSE-128, AVX-256, AVX-512, NEON-128, and VSX-128 versions on supporting CPU platforms from our testbed (Table \ref{tbl:cpu_inst_support}).

\begin{table}[b]
\centering
\renewcommand{\arraystretch}{1.25}
\begin{tabular}{|c|c|c|c|c|c|}
\hline
\rowcolor[HTML]{343434} 
{\color[HTML]{FFFFFF} \textbf{Dataset}} & {\color[HTML]{FFFFFF} \textbf{Size}} & \cellcolor[HTML]{333333}{\color[HTML]{FFFFFF} \textbf{Files}} & {\color[HTML]{FFFFFF} \textbf{Dataset Information}}                                                                                        & {\color[HTML]{FFFFFF} \textbf{XC}} & \cellcolor[HTML]{333333}{\color[HTML]{FFFFFF} \textbf{Median CDC}} \\ \hline
\textbf{DEB}                            & 40 GB                                & 65                                                            & \begin{tabular}[c]{@{}c@{}}Debian \cite{debian_org} VM Images obtained from \\ the VMware Marketplace \cite{vmware}\end{tabular}                                           & 18.98\%                            & 34.64\%                                                            \\ \hline
\textbf{DEV}                            & 230 GB                               & 100                                                           & Nightly backups of a Rust \cite{github_rust} build server                                                                                                     & 83.17\%                            & 98.05\%                                                            \\ \hline
\textbf{FLOW}                           & 8 GB                                 & 630341                                                        & C++ source code for 25 versions of TensorFlow \cite{tensorflow}                                                                                             & 90.69\%                            & 91.98\%                                                            \\ \hline
\textbf{KUBE}                           & 1.5 GB                               & 117344                                                        & Go source code for 5 versions of Kubernetes \cite{kubernetes}                                                                                                 & 64.52\%                            & 69.42\%                                                            \\ \hline
\textbf{LNX}                            & 65 GB                                & 160                                                           & Linux kernel distributions \cite{kernel_linux} in TAR format \cite{gnu_tar}                                                                                                   & 19.87\%                            & 45.62\%                                                            \\ \hline
\textbf{MAPS}                           & 981 GB                               & 15                                                            & \begin{tabular}[c]{@{}c@{}}OpenStreetMap \cite{openstreetmap} backups of Canada \\extracted using GeoFabrik \cite{geofabrik}\end{tabular}                                       & 0.10\%                                   & 68.57\%                                                                   \\ \hline
\textbf{NEWS}                           & 478 GB                               & 47                                                            & \begin{tabular}[c]{@{}c@{}}Complete snapshots of a news website across \\ 47 consecutive days in TAR \cite{gnu_tar} format\end{tabular}                   & 38.95\%                            & 73.80\%                                                            \\ \hline
\textbf{RDS}                            & 122 GB                               & 100                                                           & Redis \cite{redis} snapshots between redis-benchmark runs                                                                                               & 33.54\%                            & 92.94\%                                                            \\ \hline
\textbf{TPCC}                           & 106 GB                               & 25                                                            & 25 snapshots of a MySQL \cite{mysql} VM running TPC-C \cite{TPCCOverview}                                                                                                   & 37.39\%                            & 86.64\%                                                            \\ \hline
\textbf{WIKI}                           & 1 GB                                 & 3134                                                          & \begin{tabular}[c]{@{}c@{}}Snapshots of the largest Wikipedia article \cite{wikipediaListFilms} across \\ multiple days, chosen for extreme versioning.\end{tabular} & 1.31\%                             & 72.37\%                                                            \\ \hline
\end{tabular}
\caption{Dataset Information. Note that \textit{XC} represents the space savings achieved by fixed-size chunking with 8KB chunks while \textit{Median CDC} is the median space savings achieved by CDC algorithms with an 8KB average chunk size.}
\label{tbl:dataset_info}
\end{table}


\textbf{Datasets.} We use 10 diverse datasets to evaluate \sysname; Table \ref{tbl:dataset_info} shows their details. The datasets represent diverse workloads such as VM backups, database and map backups, web snapshots, and source code. Some datasets, such as \texttt{FLOW} and \texttt{WIKI}, are similar to those used by previous studies \cite{predict_dedup_alberta_2022}. We have publicly released the \texttt{DEB} dataset \footnote{\url{https://www.kaggle.com/datasets/sreeharshau/vm-deb-fast25}} \cite{deb_dataset}.

We note that the selected datasets have diverse characteristics. They have varying sizes, ranging from $1$ GB for \texttt{WIKI} to $981$ GB for \texttt{MAPS}. They have different file counts; datasets such as \texttt{MAPS} and \texttt{NEWS} consist of a few large files, while others, such as \texttt{FLOW} and \texttt{KUBE}, consist of a large number of small files. We include files with varying formats, such as \textit{OSM} \cite{openstreetmapFileFormats}, \textit{RDB} \cite{rdb_format}, \textit{TAR} \cite{gnu_tar}, \textit{VMDK / OVA} \cite{ovf_format}, text files, and binary files across these datasets for comprehensive coverage. 

Finally, Table \ref{tbl:dataset_info} shows the space savings achieved by using fixed-size chunking (\textit{XC}) and the median of those achieved by CDC algorithms (\textit{Median CDC}) on these datasets with 8KB chunks. By comparing \textit{XC} against \textit{Median CDC}, we note that the datasets possess varying degrees of byte-shifting. \newtext{The difference in space savings between \textit{XC} and \textit{Median CDC} in \texttt{FLOW} and \texttt{KUBE} is small (less than 6\% ), indicating a smaller number of byte-shifts.} 
\newtext{\texttt{DEV} has a moderate amount of byte-shifting, as shown by the \textasciitilde$15\%$ difference between \textit{XC} and \textit{Median CDC}. Finally, CDC algorithms achieve a median of more than $2\times$ higher space savings than \textit{XC} on \texttt{DEB}, \texttt{LNX},  \texttt{MAPS}, \texttt{NEWS}, \texttt{RDS}, \texttt{TPCC}, and \texttt{WIKI}, indicating that these data sets have a large degree of byte-shifting.}

\begin{figure}
    \centering
    \begin{subfigure}[t]{0.45\linewidth}
        \includegraphics[width=\linewidth]{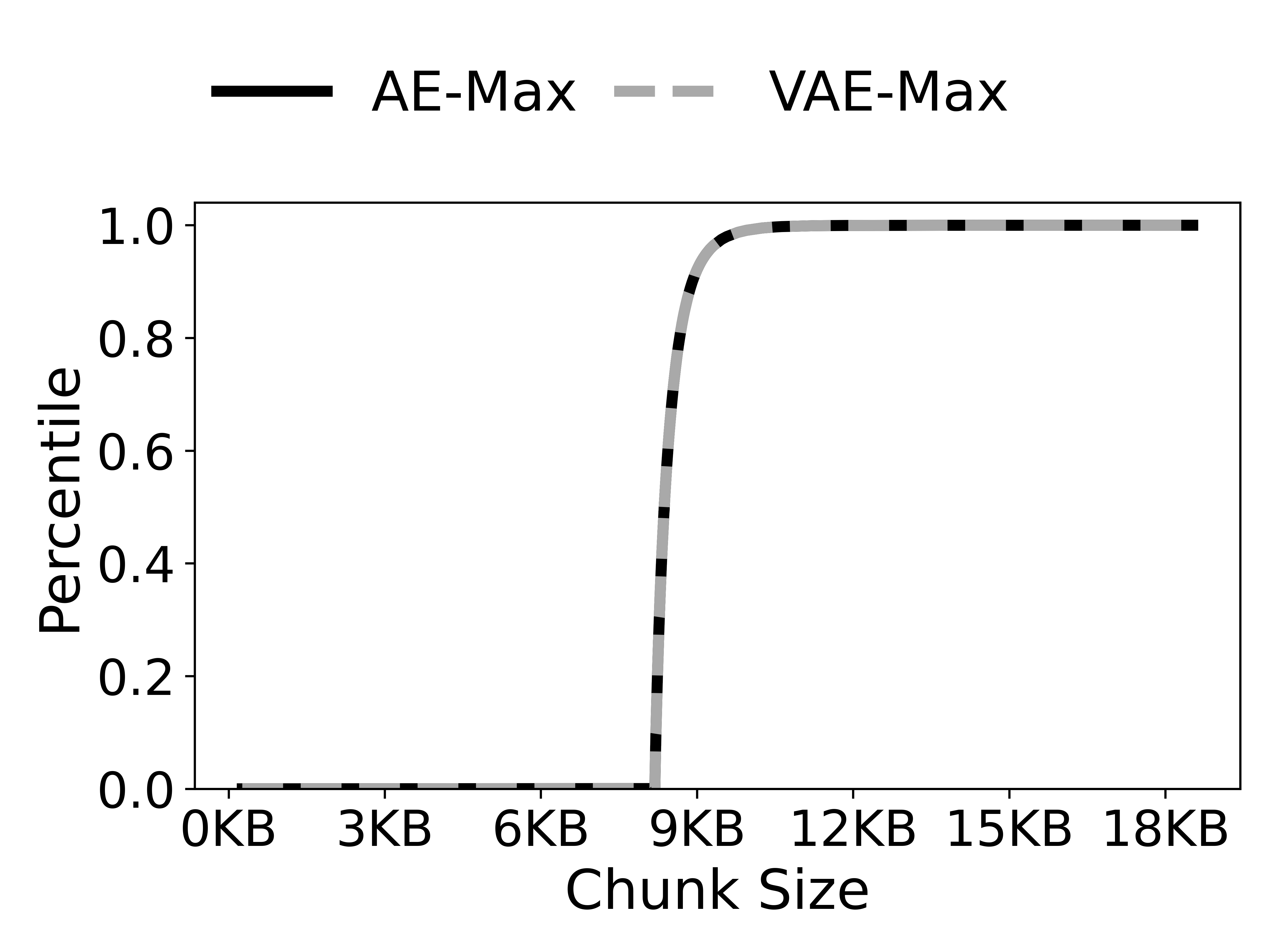}
        \caption{\textit{AE-Max} / \textit{VAE-Max}}
        \label{fig:chunksizecdf_aemax}
    \end{subfigure}
    \begin{subfigure}[t]{0.45\linewidth}
        \includegraphics[width=\linewidth]{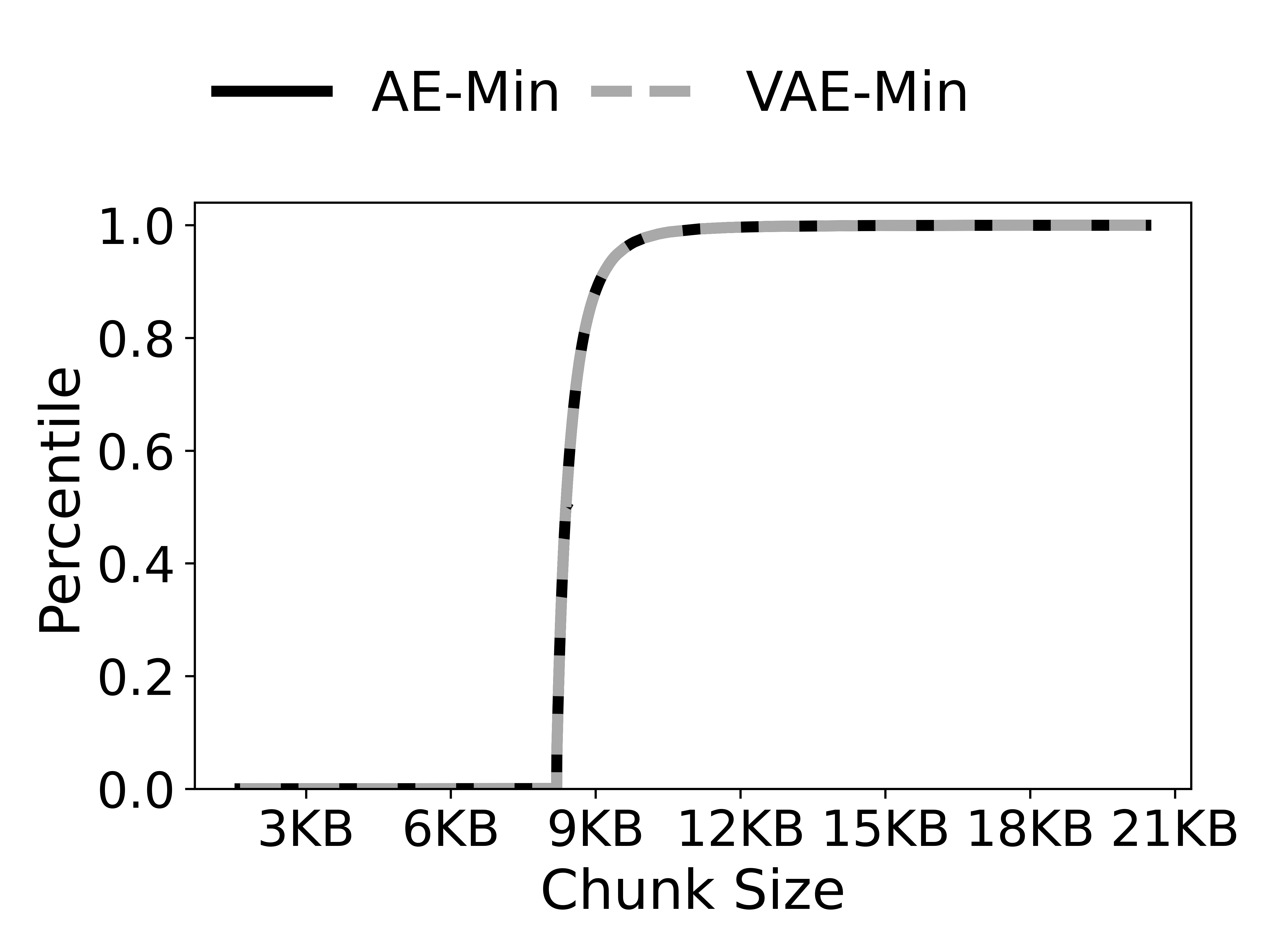}
        \caption{\textit{AE-Min} / \textit{VAE-Min}}
        \label{fig:chunksizecdf_aemin}
    \end{subfigure}
    \hfill
    \begin{subfigure}[t]{0.45\linewidth}
        \includegraphics[width=\linewidth]{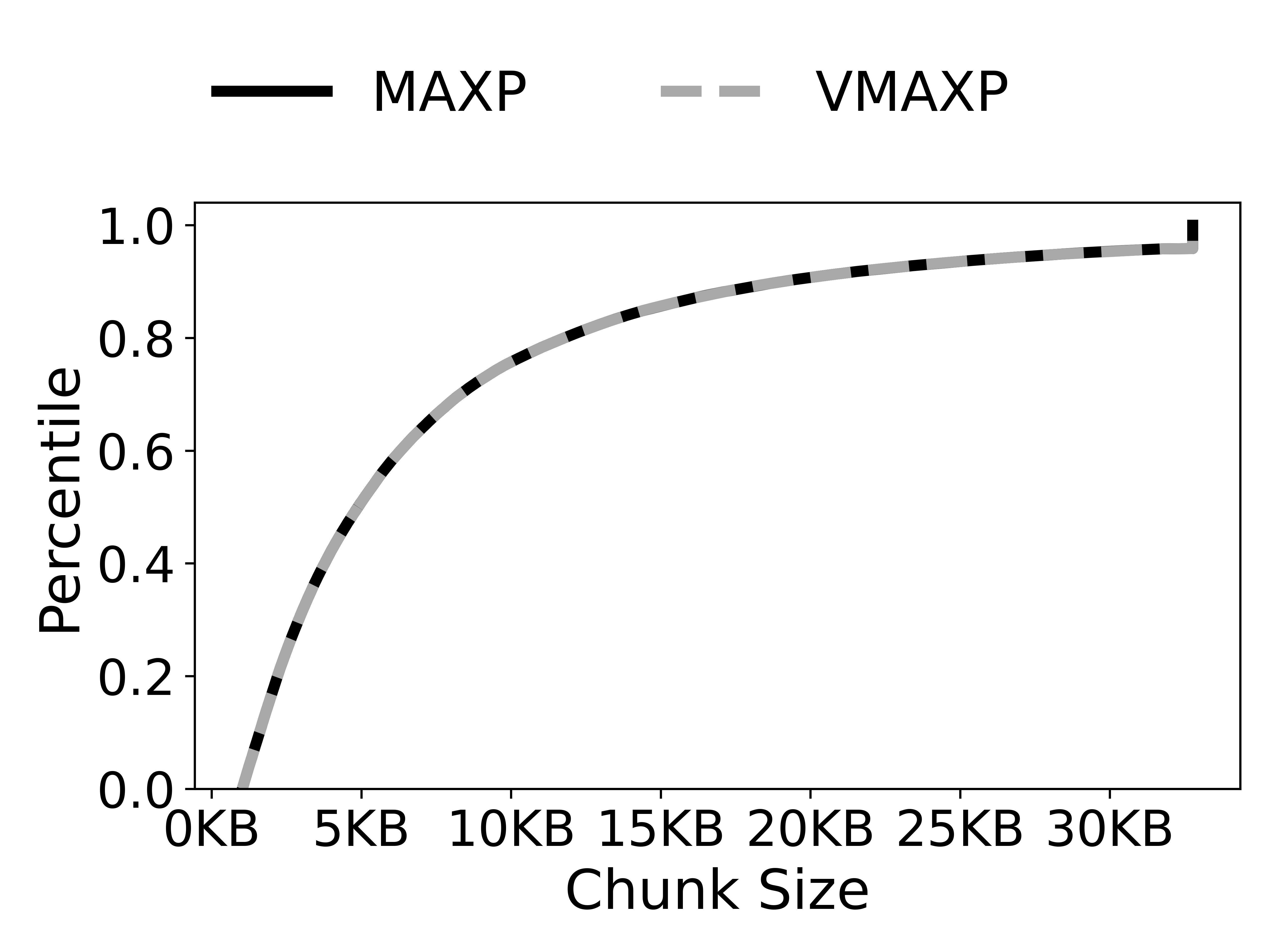}
        \caption{\textit{MAXP} / \textit{VMAXP}}
        \label{fig:chunksizecdf_maxp}
    \end{subfigure}
    \begin{subfigure}[t]{0.45\linewidth}
        \includegraphics[width=\linewidth]{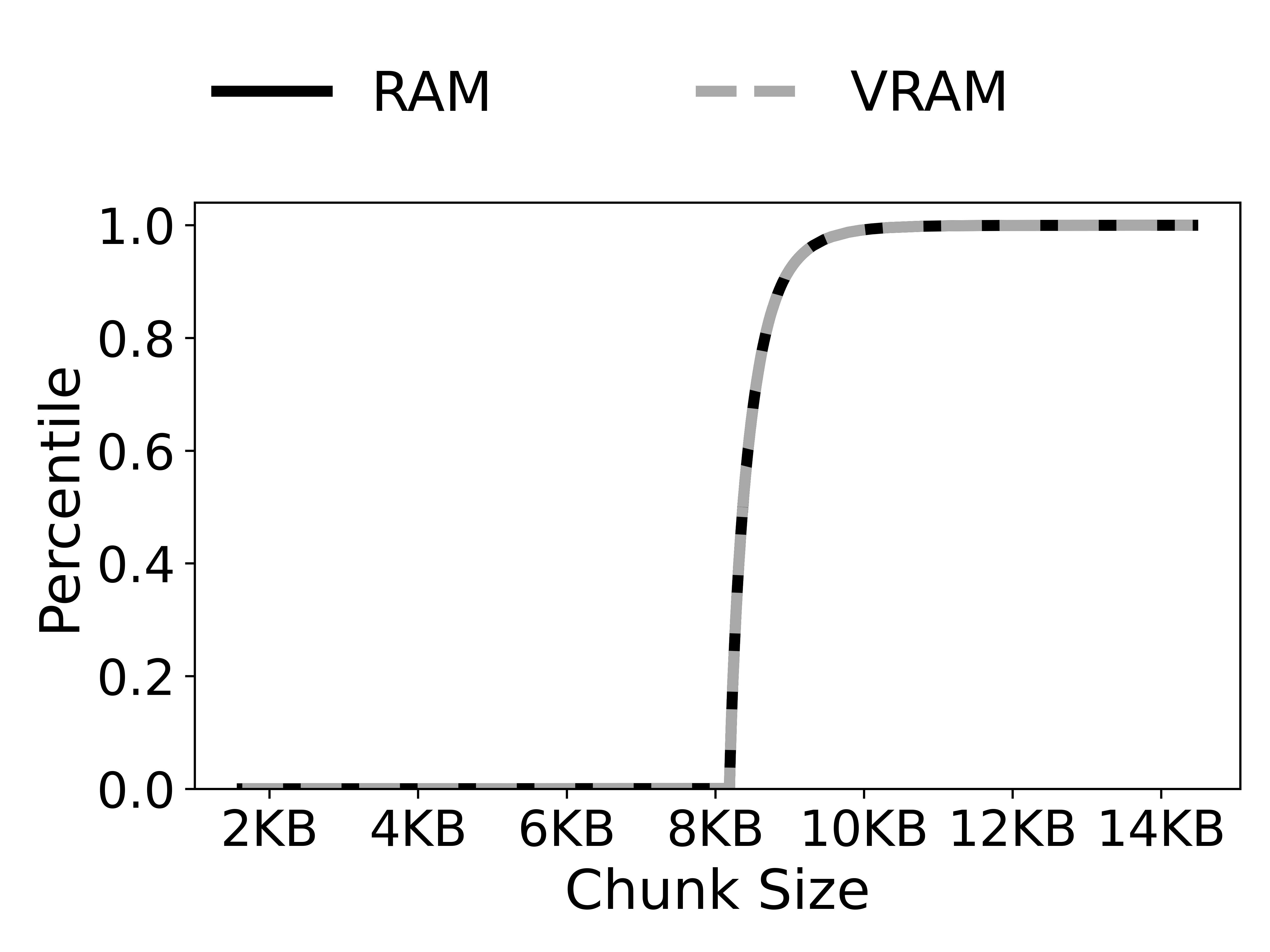}
        \caption{\textit{RAM} / \textit{VRAM}}
        \label{fig:chunksizecdf_ram}
    \end{subfigure}
    \caption{Chunk size CDFs of hashless algorithms and their AVX-512 accelerated versions on \texttt{TPCC} with an 8KB average chunk size}
    \label{fig:chunksizecdf}
\end{figure}

\textbf{Metrics.} We evaluate the space savings, chunk size distribution, and chunking throughput achieved by each alternative on all the described datasets. 

\subsection{Space Savings and Chunk Size Distributions}
\label{sec:eval_space_savings}

Figures \ref{fig:space_savings_deb} - \ref{fig:space_savings_maps} show the space savings achieved by all alternatives with 8KB chunks across datasets. We omit the results for other chunk sizes as the trends were similar.

\subsubsection{Vector-acceleration Impact} 

\begin{figure*}[t]
    \centering
    \hspace*{\fill}
     \begin{subfigure}[b]{0.7\linewidth}
        \centering
        \includegraphics[width=\linewidth]{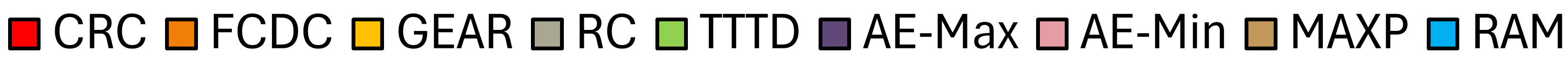}
    \end{subfigure}
    \hspace*{\fill}
    
    \begin{subfigure}[b]{0.3\linewidth}
        \includesvg[inkscapelatex=false,width=\linewidth]{figures/Space_Savings_DEB.svg}
        \caption{\texttt{DEB}}
         \label{fig:space_savings_deb}
    \end{subfigure}
    \begin{subfigure}[b]{0.3\linewidth}
        \includesvg[inkscapelatex=false,width=\linewidth]{figures/Space_Savings_DEV.svg}
        \caption{\texttt{DEV}}
         \label{fig:space_savings_dev}
    \end{subfigure}
    \begin{subfigure}[b]{0.3\linewidth}
        \includesvg[inkscapelatex=false,width=\linewidth]{figures/Space_Savings_FLOW.svg}
        \caption{\texttt{FLOW}}
        \label{fig:space_savings_flow}
    \end{subfigure} \hfill
    \begin{subfigure}[b]{0.3\linewidth}
        \includesvg[inkscapelatex=false,width=\linewidth]{figures/Space_Savings_KUBE.svg}
        \caption{\texttt{KUBE}}
        \label{fig:space_savings_kube}
    \end{subfigure}
    \begin{subfigure}[b]{0.3\linewidth}
        \includesvg[inkscapelatex=false,width=\linewidth]{figures/Space_Savings_LNX.svg}
        \caption{\texttt{LNX}}
        \label{fig:space_savings_lnx}
    \end{subfigure}
    \begin{subfigure}[b]{0.3\linewidth}
        \includesvg[inkscapelatex=false,width=\linewidth]{figures/Space_Savings_NEWS.svg}
        \caption{\texttt{NEWS}}
        \label{fig:space_savings_news}
    \end{subfigure} \hfill
    \begin{subfigure}[b]{0.3\linewidth}
        \includesvg[inkscapelatex=false,width=\linewidth]{figures/Space_Savings_RDS.svg}
        \caption{\texttt{RDS}}
        \label{fig:space_savings_rds}
    \end{subfigure}
     \begin{subfigure}[b]{0.3\linewidth}
        \includesvg[inkscapelatex=false,width=\linewidth]{figures/Space_Savings_TPCC.svg}
        \caption{\texttt{TPCC}}
        \label{fig:space_savings_tpcc}
    \end{subfigure}
    \begin{subfigure}[b]{0.3\linewidth}
        \includesvg[inkscapelatex=false,width=\linewidth]{figures/Space_Savings_WIKI.svg}
        \caption{\texttt{WIKI}}
        \label{fig:space_savings_wiki}
    \end{subfigure}
    \begin{subfigure}[b]{0.3\linewidth}
        \includesvg[inkscapelatex=false,width=\linewidth]{figures/Space_Savings_MAPS.svg}
        \caption{\texttt{MAPS}}
        \label{fig:space_savings_maps}
    \end{subfigure}
    \caption{\newtext{Space Savings with 8KB chunks. Note that the legend entries are in the same order as the plot bars.}}
    \label{fig:space_savings}
\end{figure*}

\newtext{Vector-acceleration does not impact the space savings achieved by CDC algorithms. Consequently, for clarity, we omit the space savings results for vector-accelerated algorithms from Figure \ref{fig:space_savings}.}
This aligns with the results previously observed for \textit{SS-CRC} and \textit{SS-GEAR} \cite{sscdc}. 

AVX-512 acceleration does not impact the chunks generated by hashless algorithms. We compared the \newtext{generated chunks} of vector-accelerated algorithms with their native counterparts and \newtext{verified that they were identical}. \newtext{We present only the chunk size distribution comparison in this paper due to space constraints.} 

Figure \ref{fig:chunksizecdf} shows the chunk size distributions exhibited by \textit{AE-Max}, \textit{AE-Min}, \textit{MAXP}, and \textit{RAM} compared against their AVX-512 versions accelerated with \sysname. Note that each figure is a cumulative frequency (CDF) \cite{cumulativefreq} plot. We use a target average chunk size of 8KB and the \texttt{TPCC} dataset for this experiment. The results for other datasets and chunk sizes were similar and have been omitted for clarity.


\subsubsection{Hash-based vs Hashless} 

Hashless algorithms are generally competitive with hash-based ones in space savings. The best among the hashless algorithms achieves slightly lower space savings than the best hash-based algorithm on some datasets, such as \texttt{DEB} and \texttt{NEWS} (Figures \ref{fig:space_savings_deb} and \ref{fig:space_savings_news}). On the other hand, the best hashless algorithm outperforms all hash-based algorithms on other datasets, such as \texttt{LNX} and \texttt{RDS} (Figures \ref{fig:space_savings_lnx} and \ref{fig:space_savings_rds}). Overall, the best hashless algorithms achieve space savings values within 11\% of the best hash-based ones across all datasets and chunk sizes.


\subsubsection{Hashless algorithm comparison}

The performance of the hashless algorithms depends on the dataset's characteristics and the average chunk size. For instance, \textit{RAM} achieves the highest space savings on \texttt{DEB} (Figure \ref{fig:space_savings_deb}) while \textit{MAXP} does so on \texttt{TPCC} (Figure \ref{fig:space_savings_tpcc}). This shows that \textit{accelerating all hashless algorithms is important, as the performance of each algorithm depends on the dataset's characteristics}.

Notably, \textit{AE-Min} is adversely affected by the byte-shifting pattern in \texttt{MAPS}, causing it to achieve only 8.89\% in space savings while other CDC algorithms achieve 58\%-78\%. 

Finally, while \textit{MAXP} achieves higher space savings than \textit{RAM} and both \textit{AE} variants on many datasets, the space savings difference between it and the next best hashless algorithm is small.

\newtext{\subsubsection{Differences among datasets.} Hashless algorithms perform equivalent to or better than their counterparts on virtual machine and database backups, such as \texttt{DEV}, \texttt{RDS}, and \texttt{TPCC}. Source-code datasets demonstrate mixed results, with hash-based algorithms slightly edging out hashless ones on \texttt{KUBE}, equivalence on \texttt{FLOW}, and hashless algorithms being better on \texttt{LNX}. File formats largely do not influence space savings. }

\subsection{Chunking Throughput}
\label{sec:eval_throughput}

\begin{figure}[t]
    \centering
     \begin{subfigure}[b]{0.98\linewidth}
        \includegraphics[width=\linewidth]{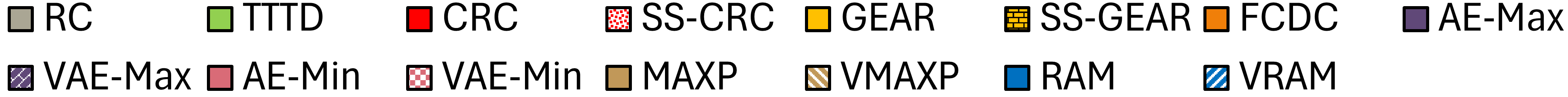}
    \end{subfigure}
    \begin{subfigure}[b]{0.48\linewidth}
        \includesvg[inkscapelatex=false,width=\linewidth]{figures/Throughput_Native_DEB.svg}
        \caption{\texttt{DEB}}
         \label{fig:chunking_throughput_native_deb}
    \end{subfigure}
    \begin{subfigure}[b]{0.48\linewidth}
        \includesvg[inkscapelatex=false,width=\linewidth]{figures/Throughput_Native_DEV.svg}
        \caption{\texttt{DEV}}
         \label{fig:chunking_throughput_native_dev}
    \end{subfigure} \hfill
    \begin{subfigure}[b]{0.45\linewidth}
        \includesvg[inkscapelatex=false,width=\linewidth]{figures/AVX_Speedups_SSCDC.svg}
        \caption{Hash-based algorithms with SS-CDC \cite{sscdc}}
        \label{fig:chunking_throughput_speedup_sscdc}
    \end{subfigure}   
    \begin{subfigure}[b]{0.45\linewidth}
        \includesvg[inkscapelatex=false,width=\linewidth]{figures/AVX_Speedups_RAM.svg}
        \caption{Hashless algorithms with \sysname}
        \label{fig:chunking_throughput_speedup_vectorcdc}
    \end{subfigure}
    \caption{Chunking Throughput with AVX-512 instructions and 8KB chunks. Note \newtext{the different scales in Figures \ref{fig:chunking_throughput_speedup_sscdc} and \ref{fig:chunking_throughput_speedup_vectorcdc}}, and that the legend entries are in the same order as the plot bars from Figures \ref{fig:chunking_throughput_native_deb} and \ref{fig:chunking_throughput_native_dev}.}
    \label{fig:chunking_throughput}
\end{figure}

Figures \ref{fig:chunking_throughput_native_deb} and \ref{fig:chunking_throughput_native_dev} show the throughput achieved by all algorithms on \texttt{DEB} and \texttt{DEV} with a chunk size of 8KB. Note that vector-accelerated algorithms are shown with patterned bars and that we have cropped the y-axis to 5 GB/s to avoid the figures being skewed by \textit{VRAM}. The results on other datasets and chunk sizes had similar trends and have been omitted for clarity. 

\subsubsection{Throughput Comparison} 

Figures \ref{fig:chunking_throughput_native_deb} and \ref{fig:chunking_throughput_native_dev} show that hashless algorithms accelerated with \sysname achieve $4\times$ to $15\times$ higher throughput than all accelerated CDC algorithms. \textit{VRAM}, the fastest accelerated hashless algorithm, achieves $8.35\times$ and $15.3\times$ higher throughput than \textit{SS-GEAR} and \textit{FastCDC}, the fastest accelerated and unaccelerated hash-based algorithms, respectively. \newtext{Additionally, \textit{VRAM} achieves $207.2\times$ higher throughput than \textit{RC}, a popular but slow hash-based CDC algorithm.}

Among unaccelerated hash-based algorithms, \textit{Gear} \cite{gear_hash}, \textit{CRC} \cite{sscdc}, and \textit{FastCDC} \cite{fastcdc} are
the fastest.
We accelerated each of these using SS-CDC \cite{sscdc}; \textit{SS-GEAR} achieves 3 $\times$ higher throughput compared to its unaccelerated version, and 
\textit{SS-CRC} achieves 2 $\times$ 
higher throughput that unaccelerated CRC. 
\newtext{We did not observe any speedup when accelerating \textit{FastCDC} \cite{fastcdc} with SS-CDC \cite{sscdc}. One of the main throughput optimizations used by \textit{FastCDC} is sub-minimum skipping (\S\ref{sec:bg_chunking}). However, as noted in \S\ref{sec:bg_sscdc}, decoupling the rolling-hash phase from the boundary identification phase eliminates the throughput benefits of minimum chunk size skipping, nullifying any speedup provided by vector-acceleration. }



\subsubsection{Vector-acceleration benefits.} Figures \ref{fig:chunking_throughput_speedup_sscdc} and \ref{fig:chunking_throughput_speedup_vectorcdc} compare the throughput benefits of accelerating hash-based and hashless algorithms with AVX-512 accelerated algorithms on \texttt{DEB} and \texttt{DEV}. 

Accelerating hash-based algorithms (Figure \ref{fig:chunking_throughput_speedup_sscdc}) using SS-CDC achieves a speedup of $2.45 - 3.32\times$. 
On the other hand, the hashless algorithms \textit{VAE-Max}, \textit{VAE-Min}, \textit{VMAXP}, and \textit{VRAM} achieve speedups of $5.1\times$, $4.43\times$, $5.36\times$, and $17.69\times$ over their respective native counterparts, achieving throughputs in the range of $6.5$ GB/s--$29.9$ GB/s. Thus, vector instructions can be leveraged far more efficiently for hashless algorithms,\textit{ proving that hashless algorithms are better candidates for vector-acceleration than their hash-based counterparts}.

\begin{figure}[t]
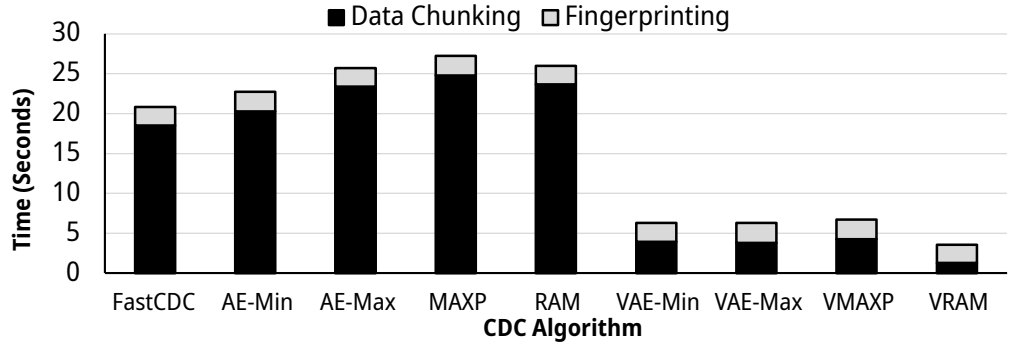

    \centering
    \begin{subfigure}{0.9\linewidth}
        \includesvg[inkscapelatex=false, width=\linewidth]{figures/evaluation_chunking_hashing_time.svg}
        \caption{\newtext{xxHash3 - 128-bit digest}}
        \label{fig:eval_chunking_hashing_time_xxhash}
    \end{subfigure}
    \begin{subfigure}{0.9\linewidth}
        \includesvg[inkscapelatex=false, width=\linewidth]{figures/evaluation_chunking_hashing_time_sha256.svg}
        \caption{\newtext{SHA256 - 256-bit digest}}
        \label{fig:eval_chunking_hashing_time_sha256}
    \end{subfigure}
    
    \caption{Time taken for Data Chunking vs Fingerprinting on \texttt{DEB} with an 8 KB average chunk size, and AVX-512 instructions for acceleration.}
    \label{fig:eval_chunking_hashing_time}
\end{figure}

Figure \ref{fig:chunking_throughput_speedup_vectorcdc} shows that \textit{VRAM} achieves higher throughputs than \textit{VAE-Max}, \textit{VAE-Min}, and \textit{VMAXP}. This is because \textit{VAE} requires multiple iterations of \textit{Range Scan} per chunk, each followed by an \textit{Extreme Byte Search}, while \textit{VRAM} only requires one iteration of each (\S\ref{sec:design}). Similarly, \textit{VMAXP} requires multiple \textit{Range Scans}, each followed by two \textit{Extreme Byte Searches}. For a given target average chunk size, the size of the \textit{Extreme Byte Search} regions in \textit{MAXP} is $70-80$\% smaller than the search region in \textit{AE}. This allows \textit{VMAXP} to achieve speeds similar to \textit{VAE-Max} and \textit{VAE-Min} despite needing an extra \textit{Extreme Byte Search}. 

Thus, \textit{RAM is inherently more vector-friendly than AE and MAXP}. However, note that \textit{VAE} and \textit{VMAXP} are still faster than every other CDC algorithm.

\subsubsection{Deduplication performance bottlenecks.} \label{sec:eval_deduplication_perf_bottlenecks} Figure \ref{fig:eval_chunking_hashing_time} shows the time taken by the chunking and hashing phases in the deduplication pipeline on \texttt{DEB} with an 8KB average chunk size. We omit the results for other datasets as they were similar. We used \newtext{two fingerprinting algorithms; xxHash3, the fastest but generates a 128-bit digest, and SHA-256, slower but offers higher collision resistance with a 256-bit digest} (\S\ref{sec:bg_dedup_bottlenecks}). We ran this experiment on the Intel Emerald Rapids machine. We use AVX-512 versions of hashless CDC algorithms, accelerated with \sysname.  

Figure \ref{fig:eval_chunking_hashing_time} shows 
\newtext{that with xxHash3 (Figure \ref{fig:eval_chunking_hashing_time_xxhash}), data chunking takes significantly longer than fingerprinting with unaccelerated algorithms. On the other hand, \textit{VAE-Min}, \textit{VAE-Max}, \textit{VMAXP}, and \textit{VRAM} show data chunking times similar to or lower than fingerprinting.} For instance, with \textit{VRAM}, data chunking takes $1.29$ seconds while fingerprinting takes $2.27$ seconds. 

\newtext{With SHA-256 (Figure \ref{fig:eval_chunking_hashing_time_sha256}), we observe that fingerprinting takes as long as data chunking with unaccelerated CDC algorithms. On the other hand, \textit{VAE-Max}, \textit{VAE-Min}, \textit{VMAXP}, and \textit{VRAM} take significantly lower time to run. 
}

\newtext{These results} show that \textit{\sysname effectively alleviates the data chunking bottleneck in the deduplication pipeline}.

   

\subsection{Throughput breakdown - Extreme Byte Search vs Range Scan}
\label{sec:eval_throughput_breakdown}

\begin{figure}[b]
    \centering
    \begin{subfigure}[t]{0.48\linewidth}
        \includesvg[inkscapelatex=false,width=\linewidth]{figures/Throughput_Breakdown_RAM.svg}
        \caption{\textit{VRAM}}
        \label{fig:chunking_throughput_breakdown_ram}
    \end{subfigure}
    \begin{subfigure}[t]{0.48\linewidth}
        \includesvg[inkscapelatex=false,width=\linewidth]{figures/Throughput_Breakdown_MAXP.svg}
        \caption{\textit{VMAXP}}
        \label{fig:chunking_throughput_breakdown_maxp}
    \end{subfigure}
    \caption{Throughput Breakdown with AVX-512 instructions. \newtext{Note that \textit{VRAM-EBS} and \textit{VMAXP-EBS} represent \textit{VRAM} and \textit{VMAXP} with only Extreme Byte Search accelerated.}}
    \label{fig:chunking_throughput_breakdown}
\end{figure}

\begin{figure}[t]
    \centering
    \begin{subfigure}[t]{0.98\linewidth}
        \centering
        \includegraphics[width=0.4\linewidth]{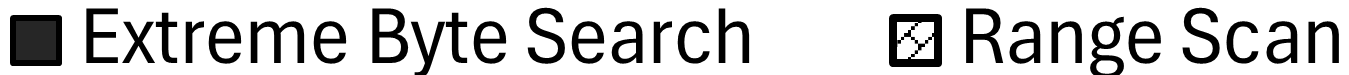}
    \end{subfigure}
    \begin{subfigure}[t]{0.4\linewidth}
            \includesvg[inkscapelatex=false,width=\linewidth]{figures/Bytes_Processed_Phasewise_DEB.svg}
        \caption{\texttt{DEB}}
        \label{fig:bytes_processed_deb}
    \end{subfigure}
     \begin{subfigure}[t]{0.4\linewidth}
        \includesvg[inkscapelatex=false,width=\linewidth]{figures/Bytes_Processed_Phasewise_LNX.svg}
        \caption{\texttt{LNX}}
        \label{fig:bytes_processed_lnx}
    \end{subfigure}
    \caption{Percentage share of bytes processed using \textit{Extreme Byte Search} and \textit{Range Scan} by hashless CDC algorithms on \texttt{DEB} and \texttt{LNX}}
    \label{fig:bytes_processed}
\end{figure}

The throughput impact of each processing pattern depends on algorithmic and dataset characteristics. Figure \ref{fig:chunking_throughput_breakdown} shows the individual impact of accelerating \textit{Extreme Byte Search} and \textit{Range Scan} using \textit{VRAM} on the \texttt{DEB} and \texttt{LNX} datasets with an 8KB chunk size. \textit{VRAM-EBS} \newtext{and \textit{VMAXP-EBS}} represent \textit{RAM} \newtext{and \textit{MAXP}} running with only \textit{Extreme Byte Search} acceleration, while \textit{VRAM-512} \newtext{and \textit{VMAXP-512}} use both accelerated patterns. 

Figure \ref{fig:chunking_throughput_breakdown_ram} shows that on \texttt{DEB}, \textit{VRAM-EBS} achieves a throughput of $18.5$ GB/s compared to \textit{RAM} at $1.7$ GB/s. Accelerating \textit{Range Scan} provides an additional speedup of $11.4$ GB/s. On the other hand on \texttt{LNX}, \textit{VRAM-EBS} only achieves $2.7$ GB/s compared to \textit{RAM} at $2$ GB/s. Accelerating \textit{Range Scan} provides an additional speedup of $27.6$ GB/s. Thus, each pattern has a balanced impact on \textit{VRAM}'s throughput on \texttt{DEB}, while \textit{Range Scan} primarily contributes to throughput on \texttt{LNX}, indicating that dataset characteristics affect the throughput breakdown.

The throughput breakdown also varies across algorithms; for instance, accelerating \textit{Extreme Byte Searches} has differing impacts on the throughputs of \textit{RAM} and \textit{MAXP}. While Figure \ref{fig:chunking_throughput_breakdown_ram} shows that \textit{VRAM-EBS} achieves significantly higher throughput than \textit{RAM} on \texttt{DEB}, Figure \ref{fig:chunking_throughput_breakdown_maxp} shows that \textit{VMAXP-EBS} only achieves small speedups over \textit{MAXP}, i.e., \textit{Range Scan} acceleration contributes more to throughput on \textit{VMAXP} than it does on \textit{VRAM}.

These results are directly tied to the number of bytes processed by the algorithms on both datasets. Figure \ref{fig:bytes_processed} shows the percentage shares of bytes processed by \textit{Extreme Byte Searches} and \textit{Range Scans}, for all hashless algorithms on \texttt{DEB} and \texttt{LNX}. As seen in Figure \ref{fig:bytes_processed_deb}, the percentage shares differ across algorithms. For instance, \textit{RAM} processes $96.70\%$ and $3.30\%$ of bytes on \texttt{DEB} with \textit{Extreme Byte Search} and \textit{Range Scan}, respectively. On the other hand, \textit{MAXP} processes $10.26\%$ and $89.74\%$ of bytes with \textit{Extreme Byte Search} and \textit{Range Scan}, respectively. Additionally, this percentage varies across datasets, as seen by the differences between Figures \ref{fig:bytes_processed_deb} and \ref{fig:bytes_processed_lnx}.


Thus, \textit{accelerating both phases using vector instructions is crucial to performance, as the impact of each phase depends on dataset and algorithmic characteristics}.


\subsection{\sysname across different vector instruction sets}
\label{sec:eval_proc_compat}

\sysname is compatible with a large range of platforms that support vector instructions such as SSE-128, AVX-256, NEON-128, and VSX-128 (\S\ref{sec:bg_vector_inst}).
This is unlike SS-CDC \cite{sscdc} which requires CPUs with \texttt{scatter/gather} instruction support.
Such CPUs are present only in a small percentage of datacenter nodes today. 

\begin{figure}[t]
    \centering
    \begin{subfigure}[b]{0.8\linewidth}
        \includegraphics[width=\linewidth]{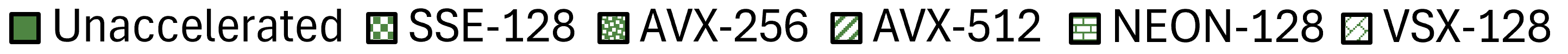}
    \end{subfigure}
    \begin{subfigure}[b]{0.48\linewidth}
        \includesvg[inkscapelatex=false,width=\linewidth]{figures/hashless_simd_amdepyc.svg}
        \caption{AMD EPYC Rome}
        \label{fig:proc_compat_amd}
    \end{subfigure}
          \begin{subfigure}[b]{0.48\linewidth}
        \includesvg[inkscapelatex=false,width=\linewidth]{figures/hashless_simd_intelemerald.svg}
        \caption{Intel Emerald Rapids}
        \label{fig:proc_compat_emerald}
    \end{subfigure}
    \begin{subfigure}[b]{0.48\linewidth}
        \includesvg[inkscapelatex=false,width=\linewidth]{figures/hashless_simd_intelskylake.svg}
        \caption{Intel Skylake}
        \label{fig:proc_compat_skylake}
    \end{subfigure}
     \begin{subfigure}[b]{0.48\linewidth}
        \includesvg[inkscapelatex=false,width=\linewidth]{figures/hashless_simd_armneon.svg}
        \caption{ARM v8 Atlas}
        \label{fig:proc_compat_arm}
    \end{subfigure}
     \begin{subfigure}[b]{0.48\linewidth}
        \includesvg[inkscapelatex=false,width=\linewidth]{figures/hashless_simd_ibmpower.svg}
        \caption{IBM Power 8}
        \label{fig:proc_compat_ibm}
    \end{subfigure}
    \caption{Accelerating hashless algorithms with \sysname across processor architectures, on \texttt{DEB} at an 8KB average chunk size. \newtext{Data labels show speedups over the respective native algorithm for the specific cases discussed in text.} Note the different y-axis scale on Figures \ref{fig:proc_compat_arm} and \ref{fig:proc_compat_ibm}.}
    \label{fig:proc_compat}
\end{figure}

While \S\ref{sec:design} discusses \sysname's design using AVX-512 instructions, the same methods can be applied to any vector instruction set that supports \texttt{VCMP}, \texttt{VMAX}, and \texttt{VMASK} operations. In this section, we evaluate \sysname's performance with other such vector instruction sets. 
We ran this experiment using the \texttt{DEB} dataset and an average chunk size of 8 KB.


\subsubsection{AMD EPYC Rome} Figure \ref{fig:proc_compat_amd} shows the throughputs achieved by hashless algorithms accelerated with \sysname on an AMD EPYC Rome machine. As shown in Table \ref{tbl:cpu_inst_support}, the AMD machine only supports SSE-128 and AVX-256 instructions. All four hashless algorithms in Figure \ref{fig:proc_compat_amd} show speedups over their native versions with both instruction sets. For instance, \textit{AE-Max} achieves $2.12\times$ and $3.43\times$ speedups with SSE-128 and AVX-256 instructions, respectively. Similar to the results in \S\ref{sec:eval_throughput} with AVX-512 instructions, \textit{RAM} achieves the highest throughput of all algorithms with both SSE-128 and AVX-256 instructions.

\subsubsection{Intel Emerald Rapids and Skylake} Figures \ref{fig:proc_compat_emerald} and \ref{fig:proc_compat_skylake} show the throughputs achieved by hashless algorithms accelerated with \sysname on Intel Emerald Rapids and Skylake machines. As shown in Table \ref{tbl:cpu_inst_support}, these machines support SSE-128, AVX-256, and AVX-512 instructions. All four hashless algorithms in Figures \ref{fig:proc_compat_emerald} and \ref{fig:proc_compat_skylake} achieve speedups over their unaccelerated versions with all instruction sets. For instance, in Figure \ref{fig:proc_compat_skylake}, \textit{AE-Max} achieves $2.29\times$, $4.91\times$, and $6.71\times$ speedups with SSE-128, AVX-256, and AVX-512 instructions, respectively. Similar to the results in \S\ref{sec:eval_throughput} with AVX-512 instructions, \textit{RAM} achieves the highest throughput of all algorithms with both SSE-128 and AVX-256 instructions.

On both platforms, all algorithms also benefit from increasing vector widths; that is, higher vector widths lead to higher throughput. The only exception is \textit{MAXP}, which does not gain as much as the other algorithms with AVX-512 instructions over AVX-256. This is related to the small window sizes used by \textit{MAXP} for its \textit{Extreme Byte Search} phases, which do not benefit from high vector widths. However, \textit{MAXP} still achieves $4.7\times$ and $5.42\times$ speedups with AVX-512 instructions over its unaccelerated version, on the Skylake and Emerald Rapids machines, respectively.

\subsubsection{ARM v8 Atlas} Figure \ref{fig:proc_compat_arm} shows the throughputs achieved by hashless algorithms accelerated with \sysname on an ARM v8 Atlas machine. As shown in Table \ref{tbl:cpu_inst_support}, the machine only supports NEON-128 instructions, an ARM equivalent to SSE-128. While the instruction set supports \textit{VMAX} and \textit{VCMP} operations, it lacks native support for \textit{VMASK} operations (\S\ref{sec:implementation}). \textit{RAM} achieves the highest throughput among all accelerated hashless algorithms at $2.91$ GB/s.

 All hashless algorithms achieve lower speedups on ARM with NEON-128 instructions, when compared to SSE-128 instructions on Intel and AMD machines. \textit{AE-Max} and \textit{AE-Min} are especially affected, achieving only $1.08\times$ and $1.05\times$ speedups, i.e. 8\% and 5\% gains with NEON-128 over their unaccelerated versions.  This is largely due to the lack of native \textit{VMASK} support, which affects \textit{Range Scans}. While our implementation uses an alternative method to achieve the same functionality, it uses four NEON-128 instructions instead of a single SSE-128 \textit{VMASK} instruction.

However, \textit{MAXP} and \textit{RAM} still achieve $1.93\times$ and $5.32\times$ speedups, respectively, showing that \sysname remains beneficial on ARM platforms with NEON-128 support. Note that these numbers are expected to improve in ARM platforms supporting SVE/SVE2 instructions \cite{arm_sve}, as they offer native \textit{VMASK} support. However, as we could not obtain such a platform for our evaluation, we leave a detailed SVE/SVE2 performance review as future work.

\subsubsection{IBM Power 8} Figure \ref{fig:proc_compat_ibm} shows the throughputs achieved by hashless algorithms accelerated with \sysname on an IBM Power 8 machine. As shown in Table \ref{tbl:cpu_inst_support}, this machine only supports VSX-128 instructions, an IBM equivalent to SSE-128. This instruction set lacks support for native \textit{VMASK} operations as well (\S\ref{sec:implementation}). \textit{RAM} achieves the highest throughput among all accelerated hashless algorithms, at $8.54$ GB/s.

Unlike ARM, all hashless algorithms exhibit considerable speedups on IBM Power 8 with \sysname. \textit{AE-Max} and \textit{AE-Min} achieve speedups of $2.92\times$ and $2.85\times$ respectively. \textit{MAXP} and \textit{RAM} achieve speedups of $7.93\times$ and $20.35\times$ respectively. Furthermore, all hashless algorithms accelerated with VSX-128 instructions achieve speedups equivalent to or greater than their counterparts accelerated with SSE-128 on Intel and AMD machines. For instance, \textit{RAM} achieves a speedup of $20.35\times$ with VSX-128 on IBM Power 8 while it achieves a speedup of $7.49\times$ and $9.94\times$ with SSE-128 on Intel Emerald Rapids and AMD EPYC Rome, respectively. 

This is because, despite the lack of native \textit{VMASK} instruction support, the alternative implementation using \texttt{vec\_bpermq} is efficient and uses just two fast VSX-128 instructions.

\subsection{Evaluation Summary}

\newtext{To summarize, the main takeaways from our evaluation are the following:}
\begin{itemize}
    \item \newtext{\sysname-based hashless algorithms achieve $15.3\times$--$207.2\times$ and $8.35\times$--$26.2\times$ higher throughput than unaccelerated and vector-accelerated hash-based algorithms respectively, showing that hashless algorithms are better candidates for vector acceleration (\S\ref{sec:eval_throughput}).}
    \item \newtext{\sysname effectively alleviates the data chunking performance bottleneck in the deduplication pipeline (\S\ref{sec:eval_deduplication_perf_bottlenecks}).}
    \item \newtext{Accelerating both \textit{Extreme Byte Search} and \textit{Range Scan} is important because their individual impact depends on dataset and algorithmic characteristics (\S\ref{sec:eval_throughput_breakdown}).}
    \item \newtext{\sysname provides benefits across different processor architectures, and is compatible with a wide range of vector instruction sets (\S\ref{sec:eval_proc_compat}).}
    \item \newtext{Accelerating hashless algorithms with \sysname does not impact their space savings and generates chunks identical to their unaccelerated counterparts.}
    \item \newtext{Hashless algorithms achieve space savings values comparable to or better than those of their hash-based counterparts on real-world datasets (\S\ref{sec:eval_space_savings}). The best performing hashless algorithm varies by dataset, showing that accelerating all of them is equally important.}
\end{itemize}
\section{Related Work}
\label{sec:related_work}



\subsubsection{Chunking optimizations.} Many efforts have been made to optimize data chunking. MUCH~\cite{much} and P-Dedupe~\cite{p_dedupe} use multiple threads to accelerate chunking. RapidCDC~\cite{rapidcdc} sometimes skips data chunking by predicting the next chunk boundary based on historical data, but requires maintaining additional metadata. Bimodal Chunking~\cite{kruus2010bimodal} initially splits the data into large chunks, and then divides duplicate adjacent chunks into smaller ones, to enhance space savings. \sysname is compatible with all of these approaches, as they build on top of existing CDC algorithms. 

Previous work~\cite{lowentropy_cloud} that analyzes the characteristics of chunks generated by CDC algorithms, is orthogonal to \sysname, as vector acceleration does not affect generated chunks.

\newtext{Our previous paper at USENIX FAST 2025 \cite{vectorcdc_fast} presented \sysname's design, but does not discuss accelerating \textit{MAXP} \cite{maxp} or \sysname's performance on varying CPU architectures. Additionally, it does not present a comprehensive evaluation of \sysname's capabilities.} 


\subsubsection{Deduplication optimizations.} Several other efforts exist to optimize the other phases of the deduplication pipeline. StoreGPU~\cite{storegpu} and GPU-Dedup \cite{gpu_dedup} accelerate chunk hash computation using GPUs. SiLo~\cite{Silo}, Sparse Indexing~\cite {sparse_indexing} and Extreme Binning~\cite{extremeBinning} optimize hash indexing. HYDRAStor \cite{hydrastor} is a distributed deduplication system that focuses on data placement. Several studies incorporate delta compression after deduplication to further compress similar but non-duplicate chunks~\cite{shilane2012wan,zou2022building,zhang2023loopdelta}.  These efforts are orthogonal to ours as we accelerate the data chunking phase. 

\subsubsection{Accelerating other storage systems.} Vector instructions have been widely used to accelerate other storage systems. MinervaFS \cite{minervafs} accelerates the computation of transform and basis functions in generalized deduplication with AVX instructions. ICID \cite{inlinededup_nvme} records memory-copy operations in a B-Tree for fine-grained deduplication, accelerating tree searches with AVX instructions. AVX-512 conflict detection instructions have been used to accelerate lightweight data compression algorithms \cite{avx_rle}. Numerous works attempt to accelerate collision-resistant hashing algorithms used across storage systems with vector instructions \cite{clhash_vectorized, murmur_vectorized}. These efforts are orthogonal to ours as we focus on using vector instructions to accelerate CDC algorithms for block-level deduplication.

\subsubsection{Secure deduplication systems.} Several efforts build end-to-end deduplication systems for encrypted data \cite{secure_dedup_survey}. They mainly target encryption schemes \cite{mle, secure_asokan} for the underlying data or focus on reducing attacks on the system \cite{side_channel_dedup, sgxdedup}. Some target specific applications, such as distributing encrypted docker images ~\cite{sun2024simenc} and encrypted videos \cite{securededup_video}. As all of these efforts layer encryption atop existing data chunking algorithms, \sysname is compatible with all these approaches.

\section {Conclusion}
\label{sec:conclusion}

We present \sysname, a methodology for accelerating content-defined chunking using vector instructions. \sysname avoids the pitfalls of previous work that accelerates CDC algorithms by choosing hashless CDC algorithms instead. \sysname accelerates these algorithms using novel \textit{tree-based search} and \textit{packed scanning} methods. Our evaluation shows that \sysname achieves $8.35\times$-$26.2\times$ higher throughput than existing vector-accelerated CDC algorithms and $15.3\times$-$207.2\times$ higher throughput than unaccelerated algorithms. We have made our code publicly available by integrating it with DedupBench \cite{dedupbench}, and published one of our datasets on Kaggle \cite{deb_dataset}.

\begin{acks}

We thank the anonymous reviewers of USENIX FAST 2025 and ACM Transactions on Storage for their feedback. We thank Lori Paniak for his technical assistance throughout the project, and Mu'men Al-Jarah for his feedback on an earlier version of this work. The research team was supported by grants from the National Cybersecurity Consortium (NCC), Natural Sciences and Engineering Research Council of Canada (NSERC), and the Ontario Research Fund's Research Excellence Program (ALLRP-561423-20, RGPIN-2025-03332, and ORF-RE012-051). The team was also supported by research grants from Acronis, Oracle Research Labs, and Rogers Communications. Sreeharsha is supported by the Cheriton Graduate Scholarship and the Ontario Graduate Scholarship.

\end{acks}

\bibliographystyle{unsrt}
\bibliography{refs}

@String{BIT = "{BIT}" }

@String{Computing = "Computing" }

@String{Computer = "{IEEE} Computer" }

@String{Springer = "Springer-Verlag" }

@inproceedings{ae,
  title={{AE}: An asymmetric extremum content defined chunking algorithm for fast and bandwidth-efficient data deduplication},
  author={Zhang, Yucheng and Jiang, Hong and Feng, Dan and Xia, Wen and Fu, Min and Huang, Fangting and Zhou, Yukun},
  booktitle={2015 IEEE Conference on Computer Communications (INFOCOM)},
  pages={1337--1345},
  year={2015},
  organization={IEEE}
}

@article{ram,
  title={A new content-defined chunking algorithm for data deduplication in cloud storage},
  author={Widodo, Ryan NS and Lim, Hyotaek and Atiquzzaman, Mohammed},
  journal={Future Generation Computer Systems},
  volume={71},
  pages={145--156},
  year={2017},
  publisher={Elsevier}
}

@article{gear_hash,
title = {Ddelta: A deduplication-inspired fast delta compression approach},
journal = {Performance Evaluation},
volume = {79},
pages = {258-272},
year = {2014},
note = {Special Issue: Performance 2014},
issn = {0166-5316},
doi = {https://doi.org/10.1016/j.peva.2014.07.016},
url = {https://www.sciencedirect.com/science/article/pii/S0166531614000790},
author = {Wen Xia and Hong Jiang and Dan Feng and Lei Tian and Min Fu and Yukun Zhou},
}

@inproceedings{lbfs,
  title={A low-bandwidth network file system},
  author={Muthitacharoen, Athicha and Chen, Benjie and Mazieres, David},
  booktitle={Proceedings of the Eighteenth ACM Symposium on Operating Systems Principles (SOSP)},
  pages={174--187},
  year={2001}
}

@inproceedings{fastcdc,
  title={{FastCDC}: A fast and efficient content-defined chunking approach for data deduplication},
  author={Xia, Wen and Zhou, Yukun and Jiang, Hong and Feng, Dan and Hua, Yu and Hu, Yuchong and Liu, Qing and Zhang, Yucheng},
  booktitle={2016 {USENIX} Annual Technical Conference ({USENIX} {ATC} 16)},
  pages={101--114},
  year={2016}
}

@inproceedings{sscdc,
  title={{S}{S}-{C}{D}{C}: A two-stage parallel content-defined chunking for deduplicating backup storage},
  author={Ni, Fan and Lin, Xing and Jiang, Song},
  booktitle={Proceedings of the 12th ACM International Conference on Systems and Storage},
  pages={86--96},
  year={2019}
}

@article{tttd,
  title={A framework for analyzing and improving content-based chunking algorithms},
  author={Eshghi, Kave and Tang, Hsiu Khuern},
  journal={Hewlett-Packard Labs Technical Report TR},
  volume={30},
  number={2005},
  year={2005}
}

@inproceedings {cloudlab_paper,
author = {Dmitry Duplyakin and Robert Ricci and Aleksander Maricq and Gary Wong and Jonathon Duerig and Eric Eide and Leigh Stoller and Mike Hibler and David Johnson and Kirk Webb and Aditya Akella and Kuangching Wang and Glenn Ricart and Larry Landweber and Chip Elliott and Michael Zink and Emmanuel Cecchet and Snigdhaswin Kar and Prabodh Mishra},
title = {The Design and Operation of {CloudLab}},
booktitle = {2019 USENIX Annual Technical Conference (USENIX ATC 19)},
year = {2019},
isbn = {978-1-939133-03-8},
address = {Renton, WA},
pages = {1--14},
url = {https://www.usenix.org/conference/atc19/presentation/duplyakin},
publisher = {USENIX Association},
month = jul,
}

@article{maxp,
  title={Content-dependent chunking for differential compression, the local maximum approach},
  author={Bj{\o}rner, Nikolaj and Blass, Andreas and Gurevich, Yuri},
  journal={Journal of Computer and System Sciences},
  volume={76},
  number={3-4},
  pages={154--203},
  year={2010},
  publisher={Elsevier}
}

@article{dedup_intro,
  title={A study of practical deduplication},
  author={Meyer, Dutch T and Bolosky, William J},
  journal={ACM Transactions on Storage (ToS)},
  volume={7},
  number={4},
  pages={1--20},
  year={2012},
  publisher={ACM New York, NY, USA}
}

@inproceedings{sha256,
  title={{A comparative study of Message Digest 5 (MD5) and SHA256 algorithm}},
  author={Rachmawati, Dian and Tarigan, JT and Ginting, ABC},
  booktitle={Journal of Physics: Conference Series},
  volume={978},
  pages={012116},
  year={2018},
  organization={IOP Publishing}
}

@article{software_defined_storage,
  title={{Software Defined Storage}},
  author={Carlson, Mark and Yoder, Alan and Schoeb, Leah and Deel, Don and Pratt, Carlos and Lionetti, Chris and Voigt, Doug},
  journal={Storage Networking Industry Association Working Draft},
  pages={20--24},
  year={2014}
}

@article{raid,
  title={{RAID: High-performance, reliable secondary storage}},
  author={Chen, Peter M and Lee, Edward K and Gibson, Garth A and Katz, Randy H and Patterson, David A},
  journal={ACM Computing Surveys (CSUR)},
  volume={26},
  number={2},
  pages={145--185},
  year={1994},
  publisher={ACM New York, NY, USA}
}

@inproceedings{hadoop,
  title={{The Hadoop distributed file system}},
  author={Shvachko, Konstantin and Kuang, Hairong and Radia, Sanjay and Chansler, Robert},
  booktitle={2010 IEEE 26th Symposium on Mass Storage Systems and Technologies (MSST)},
  pages={1--10},
  year={2010},
  organization={Ieee}
}

@inproceedings{ceph,
  title={Ceph: A scalable, high-performance distributed file system},
  author={Weil, Sage and Brandt, Scott A and Miller, Ethan L and Long, Darrell DE and Maltzahn, Carlos},
  booktitle={Proceedings of the 7th Conference on Operating Systems Design and Implementation (OSDI'06)},
  pages={307--320},
  year={2006}
}

@article{memcached,
  title={Distributed caching with memcached},
  author={Fitzpatrick, Brad},
  journal={Linux Journal},
  volume={2004},
  number={124},
  pages={5},
  year={2004},
  publisher={Belltown Media Houston, TX}
}

@inproceedings{facebook_tao,
  title={{TAO}: {Facebook’s} distributed data store for the social graph},
  author={Bronson, Nathan and Amsden, Zach and Cabrera, George and Chakka, Prasad and Dimov, Peter and Ding, Hui and Ferris, Jack and Giardullo, Anthony and Kulkarni, Sachin and Li, Harry and others},
  booktitle={2013 USENIX Annual Technical Conference (USENIX ATC 13)},
  pages={49--60},
  year={2013}
}

@article{comprehensive_dedup,
  title={A comprehensive study of the past, present, and future of data deduplication},
  author={Xia, Wen and Jiang, Hong and Feng, Dan and Douglis, Fred and Shilane, Philip and Hua, Yu and Fu, Min and Zhang, Yucheng and Zhou, Yukun},
  journal={Proceedings of the IEEE},
  volume={104},
  number={9},
  pages={1681--1710},
  year={2016},
  publisher={IEEE}
}

@inproceedings{primary_dedup,
  title={{Primary Data Deduplication — Large} scale study and system design},
  author={El-Shimi, Ahmed and Kalach, Ran and Kumar, Ankit and Ottean, Adi and Li, Jin and Sengupta, Sudipta},
  booktitle={2012 USENIX Annual Technical Conference (USENIX ATC 12)},
  pages={285--296},
  year={2012}
}

@inproceedings{backup_workload_charz,
  title={Characteristics of backup workloads in production systems.},
  author={Wallace, Grant and Douglis, Fred and Qian, Hangwei and Shilane, Philip and Smaldone, Stephen and Chamness, Mark and Hsu, Windsor},
  booktitle={USENIX Conference on File and Storage Technologies (FAST)},
  volume={12},
  pages={4--4},
  year={2012}
}

@inproceedings{dedupbench,
  title={{DedupBench: A Benchmarking Tool for Data Chunking Techniques}},
  author={Liu, Alan and Baba, Abdelrahman and Udayashankar, Sreeharsha and Al-Kiswany, Samer},
  booktitle={2023 IEEE Canadian Conference on Electrical and Computer Engineering (CCECE)},
  pages={469--474},
  year={2023},
  organization={IEEE}
}

@inproceedings{data_security,
  title={Data security and privacy protection issues in cloud computing},
  author={Chen, Deyan and Zhao, Hong},
  booktitle={2012 International Conference on Computer Science and Electronics Engineering},
  volume={1},
  pages={647--651},
  year={2012},
  organization={IEEE}
}

@article{vector_instruction_sets,
  title={Vector instruction set support for conditional operations},
  author={Smith, James E and Faanes, Greg and Sugumar, Rabin},
  journal={ACM SIGARCH Computer Architecture News},
  volume={28},
  number={2},
  pages={260--269},
  year={2000},
  publisher={ACM New York, NY, USA}
}

@article{avx_matrixmul,
  title={{Effective implementation of matrix--vector multiplication on Intel's AVX multicore processor}},
  author={Hassan, Somaia A and Mahmoud, Mountasser MM and Hemeida, AM and Saber, Mahmoud A},
  journal={Computer Languages, Systems \& Structures},
  volume={51},
  pages={158--175},
  year={2018},
  publisher={Elsevier}
}

@article{avx_quicksort,
  title={{Fast quicksort implementation using AVX instructions}},
  author={Gueron, Shay and Krasnov, Vlad},
  journal={The Computer Journal},
  volume={59},
  number={1},
  pages={83--90},
  year={2016},
  publisher={Oxford University Press}
}

@inproceedings{avx_multimedia,
  title={{Vector LLVA}: a virtual vector instruction set for media processing},
  author={Bocchino Jr, Robert L and Adve, Vikram S},
  booktitle={Proceedings of the 2nd International Conference on Virtual Execution Environments},
  pages={46--56},
  year={2006}
}

@misc{statista2024,
  author    = {Statista},
  title     = {Worldwide data created from 2010 to 2025},
  year      = {2024},
  url       = {https://www.statista.com/statistics/871513/worldwide-data-created/},
}

@misc{vmware, 
    author={{V}{M}{W}are},    
    title={{V}{M}{W}are Marketplace},
    howpublished = {\url{https://marketplace.cloud.vmware.com/services}},
    year={2023},
}

@misc{github_rust,
	author = {Rust},
	title = {{G}it{H}ub - rust-lang/rust: {E}mpowering everyone to build reliable and efficient software.},
	howpublished = {\url{https://github.com/rust-lang/rust}},
	year = {2023},
	}

@misc{kernel_linux,
	author = {Linux},
	title = {{T}he {L}inux {K}ernel {A}rchives},
	howpublished = {\url{https://www.kernel.org/}},
	year = {2023},
	}

@misc{redis,
	author = {Redis},
	title = {{R}edis},
	howpublished = {\url{https://redis.io/}},
	year = {2023},
	}

@misc{mysql,
	author = {MySQL},
	title = {{M}y{S}{Q}{L}},
	howpublished = {\url{https://www.mysql.com/}},
	year = {2023},
}

@misc{TPCCOverview,
	author = {Transaction Processing Council},
	title = {{T}{P}{C}-{C} {O}verview},
	howpublished = {\url{https://www.tpc.org/tpcc/detail5.asp}},
	year= {2023},
}

@inproceedings{sparse_indexing,
  title={{Sparse indexing: Large scale, inline deduplication using sampling and locality.}},
  author={Lillibridge, Mark and Eshghi, Kave and Bhagwat, Deepavali and Deolalikar, Vinay and Trezis, Greg and Camble, Peter},
  booktitle={USENIX Conference on File and Storage Technologies (FAST)},
  volume={9},
  pages={111--123},
  year={2009}
}

@inproceedings{hydrastor,
  title={{HYDRAstor: A scalable secondary storage.}},
  author={Dubnicki, Cezary and Gryz, Leszek and Heldt, Lukasz and Kaczmarczyk, Michal and Kilian, Wojciech and Strzelczak, Przemyslaw and Szczepkowski, Jerzy and Ungureanu, Cristian and Welnicki, Michal},
  booktitle={USENIX Conference on File and Storage Technologies (FAST)},
  volume={9},
  pages={197--210},
  year={2009}
}

@inproceedings{storegpu,
author = {Al-Kiswany, Samer and Gharaibeh, Abdullah and Santos-Neto, Elizeu and Yuan, George and Ripeanu, Matei},
title = {Store{G}{P}{U}: {E}xploiting {G}raphics {P}rocessing {U}nits to {A}ccelerate {D}istributed {S}torage {S}ystems},
year = {2008},
isbn = {9781595939975},
publisher = {Association for Computing Machinery},
address = {New York, NY, USA},
url = {https://doi.org/10.1145/1383422.1383443},
doi = {10.1145/1383422.1383443},
booktitle = {Proceedings of the 17th International Symposium on High Performance Distributed Computing},
pages = {165–174},
numpages = {10},
location = {Boston, MA, USA},
series = {HPDC '08}
}

@ARTICLE{MUCH,
  author={Won, Youjip and Lim, Kyeongyeol and Min, Jaehong},
  journal={IEEE Transactions on Computers}, 
  title={{MUCH: Multithreaded Content-Based File Chunking}}, 
  year={2015},
  volume={64},
  number={5},
  pages={1375-1388},
  doi={10.1109/TC.2014.2322600}}

@ARTICLE{silo,
  author={Xia, Wen and Jiang, Hong and Feng, Dan and Hua, Yu},
  journal={IEEE Transactions on Computers}, 
  title={{Similarity and Locality Based Indexing for High Performance Data Deduplication}}, 
  year={2015},
  volume={64},
  number={4},
  pages={1162-1176},
  doi={10.1109/TC.2014.2308181}
}

@inproceedings{rapidcdc,
author = {Ni, Fan and Jiang, Song},
title = {{RapidCDC: Leveraging Duplicate Locality to Accelerate Chunking in CDC-Based Deduplication Systems}},
year = {2019},
isbn = {9781450369732},
publisher = {Association for Computing Machinery},
address = {New York, NY, USA},
url = {https://doi.org/10.1145/3357223.3362731},
doi = {10.1145/3357223.3362731},
booktitle = {Proceedings of the ACM Symposium on Cloud Computing},
pages = {220–232},
numpages = {13},
location = {Santa Cruz, CA, USA},
series = {SoCC '19}
}

@INPROCEEDINGS{p_dedupe,
  author={Xia, Wen and Jiang, Hong and Feng, Dan and Tian, Lei and Fu, Min and Wang, Zhongtao},
  booktitle={2012 IEEE Seventh International Conference on Networking, Architecture, and Storage}, 
  title={{P-Dedupe: Exploiting Parallelism in Data Deduplication System}}, 
  year={2012},
  volume={},
  number={},
  pages={338-347},
  doi={10.1109/NAS.2012.46}}

@article{secure_dedup_survey,
author = {Shin, Youngjoo and Koo, Dongyoung and Hur, Junbeom},
title = {{A Survey of Secure Data Deduplication Schemes for Cloud Storage Systems}},
year = {2017},
issue_date = {December 2017},
publisher = {Association for Computing Machinery},
address = {New York, NY, USA},
volume = {49},
number = {4},
issn = {0360-0300},
url = {https://doi.org/10.1145/3017428},
doi = {10.1145/3017428},
journal = {ACM Computing Surveys},
month = {Jan},
articleno = {74},
numpages = {38}
}

@inproceedings{mle,
  title={Message-locked encryption and secure deduplication},
  author={Bellare, Mihir and Keelveedhi, Sriram and Ristenpart, Thomas},
  booktitle={Annual International Conference on the Theory and Applications of Cryptographic Techniques},
  pages={296--312},
  year={2013},
  organization={Springer}
}

@inproceedings{secure_asokan,
author = {Liu, Jian and Asokan, N. and Pinkas, Benny},
title = {{Secure Deduplication of Encrypted Data without Additional Independent Servers}},
year = {2015},
isbn = {9781450338325},
publisher = {Association for Computing Machinery},
address = {New York, NY, USA},
url = {https://doi.org/10.1145/2810103.2813623},
doi = {10.1145/2810103.2813623},
booktitle = {Proceedings of the 22nd ACM SIGSAC Conference on Computer and Communications Security},
pages = {874–885},
numpages = {12},
location = {Denver, Colorado, USA},
series = {CCS '15}
}

@ARTICLE{side_channel_dedup,
  author={Harnik, Danny and Pinkas, Benny and Shulman-Peleg, Alexandra},
  journal={IEEE Security and Privacy}, 
  title={{Side Channels in Cloud Services: Deduplication in Cloud Storage}}, 
  year={2010},
  volume={8},
  number={6},
  pages={40-47},
  doi={10.1109/MSP.2010.187}}

@inproceedings{scatter_gather_perf,
author = {Lavin, Patrick and Young, Jeffrey and Vuduc, Richard and Riedy, Jason and Vose, Aaron and Ernst, Daniel},
title = {{Evaluating Gather and Scatter Performance on CPUs and GPUs}},
year = {2021},
isbn = {9781450388993},
publisher = {Association for Computing Machinery},
address = {New York, NY, USA},
url = {https://doi.org/10.1145/3422575.3422794},
doi = {10.1145/3422575.3422794},
booktitle = {Proceedings of the International Symposium on Memory Systems},
pages = {209–222},
numpages = {14},
location = {Washington, DC, USA},
series = {MEMSYS '20}
}

@inproceedings{tunable_encrypted_dedup,
  title={Balancing storage efficiency and data confidentiality with tunable encrypted deduplication},
  author={Li, Jingwei and Yang, Zuoru and Ren, Yanjing and Lee, Patrick PC and Zhang, Xiaosong},
  booktitle={Proceedings of the Fifteenth European Conference on Computer Systems},
  pages={1--15},
  year={2020}
}

@article{persistent_memory_dedup,
  title={{Optimizing the Performance of Consistency-Aware Deduplication Using Persistent Memory}},
  author={Song, Chunlin and Chen, Xianzhang and Liu, Duo and Li, Jiali and Tan, Yujuan and Ren, Ao},
  journal={IEEE Transactions on Computer-Aided Design of Integrated Circuits and Systems},
  year={2023},
  publisher={IEEE}
}

@inproceedings{gpu_dedup,
  title={Accelerating the cloud backup using {GPU} based data deduplication},
  author={Suttisirikul, Kiatchumpol and Uthayopas, Putchong},
  booktitle={2012 IEEE 18th International Conference on Parallel and Distributed Systems},
  pages={766--769},
  year={2012},
  organization={IEEE}
}

@INPROCEEDINGS{lowentropy_cloud,
  author={Jarah, Mu'men Al and Udayashankar, Sreeharsha and Baba, Abdelrahman and Al-Kiswany, Samer},
  booktitle={2024 IEEE 17th International Conference on Cloud Computing (CLOUD)}, 
  title={{The Impact of Low-Entropy on Chunking Techniques for Data Deduplication}}, 
  year={2024},
  volume={},
  number={},
  pages={134-140},
  doi={10.1109/CLOUD62652.2024.00025}}

@misc{deb_dataset,
	title={{VM Images for Deduplication}},
	howpublished = {\url{https://www.kaggle.com/dsv/10561721}},
	DOI={10.34740/KAGGLE/DSV/10561721},
	publisher={Kaggle},
	author={Sreeharsha Udayashankar and Abdelrahman Baba and Samer Al-Kiswany},
	year={2025}
}

@article{documentduplication_2011,
  title={Document duplication: How users (struggle to) manage file copies and versions},
  author={Henderson, Sarah},
  journal={Proceedings of the American Society for Information Science and Technology},
  volume={48},
  number={1},
  pages={1--10},
  year={2011},
  publisher={Wiley Online Library}
}

@article{wandeltacomp,
  title={Wan-optimized replication of backup datasets using stream-informed delta compression},
  author={Shilane, Phlip and Huang, Mark and Wallace, Grant and Hsu, Windsor},
  journal={ACM Transactions on Storage (ToS)},
  volume={8},
  number={4},
  pages={1--26},
  year={2012},
  publisher={ACM New York, NY, USA}
}

@inproceedings{venti,
  title={Venti: A new approach to archival data storage},
  author={Quinlan, Sean and Dorward, Sean},
  booktitle={USENIX Conference on File and Storage Technologies},
  year={2002}
}

@article{oceanstore,
  title={Oceanstore: An architecture for global-scale persistent storage},
  author={Kubiatowicz, John and Bindel, David and Chen, Yan and Czerwinski, Steven and Eaton, Patrick and Geels, Dennis and Gummadi, Ramakrishna and Rhea, Sean and Weatherspoon, Hakim and Weimer, Westley and others},
  journal={ACM SIGOPS Operating Systems Review},
  volume={34},
  number={5},
  pages={190--201},
  year={2000},
  publisher={ACM New York, NY, USA}
}

@misc{intelSIMDInstructions,
	author = {Intel},
	title = {{I}ntel® {I}nstruction {S}et {E}xtensions {T}echnology},
	howpublished = {\url{https://www.intel.com/content/www/us/en/support/articles/000005779/processors.html}},
	year = {},
}

@misc{AMD_SSE128,
	author = {Advanced Micro Devices},
	title = {{R}evision {G}uide for {AMD A}thlon 64 and {AMD O}pteron$^{TM}$ {P}rocessors},
	howpublished = {\url{https://www.amd.com/content/dam/amd/en/documents/archived-tech-docs/revision-guides/25759.pdf}},
	year = {2003},
}

@misc{amdzen4_wikichip,
	author = {WikiChip},
	title = {{Z}en 4 - {M}icroarchitectures - {A}{M}{D}},
	howpublished = {\url{https://en.wikichip.org/wiki/amd/microarchitectures/zen_4}},
	year = {2022},
}

@misc{skylake_wikichip,
	author = {WikiChip},
	title = {{Skylake Server} - {M}icroarchitectures - {Intel}},
	howpublished = {\url{https://en.wikichip.org/wiki/intel/microarchitectures/skylake_(server)}},
	year = {2017},
}

@misc{intelIntrinsicsGuide,
	author = {Intel},
	title = {{I}ntel® {I}ntrinsics {G}uide},
	howpublished = {\url{https://www.intel.com/content/www/us/en/docs/intrinsics-guide/index.html}},
	year = {2024},
}

@article{avx_fluidmech,
  title={Performance analysis of {SSE} and {AVX} instructions in multi-core {CPU}s and {GPU} computing on {FDTD} scheme for solid and fluid vibration problems},
  author={Franc{\'e}s, Jorge and Bleda, Sergio and M{\'a}rquez, Andr{\'e}s and Neipp, Cristian and Gallego, Sergi and Otero, Beatriz and Bel{\'e}ndez, Augusto},
  journal={The Journal of Supercomputing},
  volume={70},
  pages={514--526},
  year={2014},
  publisher={Springer}
}

@INPROCEEDINGS{predict_dedup_alberta_2022,
  author={Randall, Owen and Lu, Paul},
  booktitle={2022 IEEE International Conference on Big Data (Big Data)}, 
  title={Predicting Deduplication Performance: An Analytical Model and Empirical Evaluation}, 
  year={2022},
  volume={},
  number={},
  pages={319-328},
  doi={10.1109/BigData55660.2022.10020871}}

@misc{wikipediaListFilms,
	author = {Wikipedia},
	title = {{L}ist of films based on actual events},
	howpublished = {\url{https://en.wikipedia.org/wiki/List_of_films_based_on_actual_events}},
	year = {2022},
}

@article{tensorflow,
  title={Deep learning with tensorflow: A review},
  author={Pang, Bo and Nijkamp, Erik and Wu, Ying Nian},
  journal={Journal of Educational and Behavioral Statistics},
  volume={45},
  number={2},
  pages={227--248},
  year={2020},
  publisher={SAGE Publications Sage CA: Los Angeles, CA}
}

@book{kubernetes,
  title={Kubernetes in action},
  author={Luksa, Marko},
  year={2017},
  publisher={Simon and Schuster}
}

@misc{gnu_tar,
	author = {GNU},
	title = {{G}{N}{U} tar 1.35: {B}asic {T}ar {F}ormat},
	howpublished = {\url{https://www.gnu.org/software/tar/manual/html_section/Standard.html}},
	year = {2023},
}

@misc{ovf_format,
	author = {DMTF},
	title = {Open Virtualization Format White Paper},
	howpublished = {\url{https://www.dmtf.org/sites/default/files/standards/documents/DSP2017_1.0.0.pdf}},
	year = {2009},
}

@misc{debian_org,
	author = {Debian},
	title = {{D}ebian -- {T}he {U}niversal {O}perating {S}ystem},
	howpublished = {\url{https://www.debian.org/}},
	year = {2025},
}

@misc{geofabrik,
	author = {GeoFabrik},
	title = {{G}{E}{O}{F}{A}{B}{R}{I}{K}},
	howpublished = {\url{https://www.geofabrik.de/}},
	year = {2025},
}

@ARTICLE{openstreetmap,
  author={Haklay, Mordechai and Weber, Patrick},
  journal={IEEE Pervasive Computing}, 
  title={{OpenStreetMap: User-Generated Street Maps}}, 
  year={2008},
  doi={10.1109/MPRV.2008.80}}

@misc{openstreetmapFileFormats,
	author = {OpenStreetMap},
	title = {{O}{S}{M} file formats - {O}pen{S}treet{M}ap {W}iki},
	howpublished = {\url{https://wiki.openstreetmap.org/wiki/OSM_file_formats}},
	year = {2025},
}

@misc{armNeonOverview,
	author = {ARM},
	title = {{ARM} {NEON} {A}rchitecture {O}verview},
	howpublished = {\url{https://developer.arm.com/documentation/dht0002/a/Introducing-NEON/NEON-architecture-overview/NEON-instructions}},
	year = {2013},
}

@misc{armPortingVector,
	author = {Kutenin, Danila},
	title = {{P}orting x86 vector bitmask optimizations to {A}rm {N}{E}{O}{N}},
	howpublished = {\url{https://community.arm.com/arm-community-blogs/b/servers-and-cloud-computing-blog/posts/porting-x86-vector-bitmask-optimizations-to-arm-neon}},
	year = {2022},
}

@article{arm_sve,
  title={{ARM SVE Unleashed: Performance and Insights Across HPC Applications on Nvidia Grace}},
  author={Shi, Ruimin and Schieffer, Gabin and Gokhale, Maya and Lin, Pei-Hung and Patel, Hiren and Peng, Ivy},
  journal={European Conference on Parallel Processing},
  year={2025}
}

@inproceedings{extremeBinning,
  title={Extreme binning: Scalable, parallel deduplication for chunk-based file backup},
  author={Bhagwat, Deepavali and Eshghi, Kave and Long, Darrell DE and Lillibridge, Mark},
  booktitle={2009 IEEE International Symposium on Modeling, Analysis \& Simulation of Computer and Telecommunication Systems},
  pages={1--9},
  year={2009},
  organization={IEEE}
}

@inproceedings{kruus2010bimodal,
  title={Bimodal content defined chunking for backup streams.},
  author={Kruus, Erik and Ungureanu, Cristian and Dubnicki, Cezary},
  booktitle={Fast},
  pages={239--252},
  year={2010}
}

@article{cumulativefreq,
  title={{C}umulative {F}requency {F}unctions},
  author={Burr, Irving W},
  journal={The Annals of Mathematical Statistics},
  volume={13},
  number={2},
  pages={215--232},
  year={1942},
  publisher={JSTOR}
}

@ARTICLE{ibm_power8,
  author={Sinharoy, B. and Van Norstrand, J. A. and Eickemeyer, R. J. and Le, H. Q. and Leenstra, J. and Nguyen, D. Q. and Konigsburg, B. and Ward, K. and Brown, M. D. and Moreira, J. E. and Levitan, D. and Tung, S. and Hrusecky, D. and Bishop, J. W. and Gschwind, M. and Boersma, M. and Kroener, M. and Kaltenbach, M. and Karkhanis, T. and Fernsler, K. M.},
  journal={IBM Journal of Research and Development}, 
  title={{IBM POWER8 processor core microarchitecture}}, 
  year={2015},
  volume={59},
  number={1},
  pages={2:1-2:21},
  doi={10.1147/JRD.2014.2376112}}

@inproceedings{sun2024simenc,
  title={$\{$SimEnc$\}$: A $\{$High-Performance$\}$$\{$Similarity-Preserving$\}$ Encryption Approach for Deduplication of Encrypted Docker Images},
  author={Sun, Tong and Jiang, Bowen and Li, Borui and Lv, Jiamei and Gao, Yi and Dong, Wei},
  booktitle={2024 USENIX Annual Technical Conference (USENIX ATC 24)},
  pages={615--630},
  year={2024}
}

@article{shilane2012wan,
  title={{WAN}-optimized replication of backup datasets using stream-informed delta compression},
  author={Shilane, Phlip and Huang, Mark and Wallace, Grant and Hsu, Windsor},
  journal={ACM Transactions on Storage (ToS)},
  volume={8},
  number={4},
  pages={1--26},
  year={2012},
  publisher={ACM New York, NY, USA}
}

@inproceedings{zou2022building,
  title={Building a high-performance fine-grained deduplication framework for backup storage with high deduplication ratio},
  author={Zou, Xiangyu and Xia, Wen and Shilane, Philip and Zhang, Haijun and Wang, Xuan},
  booktitle={2022 USENIX Annual Technical Conference (USENIX ATC 22)},
  pages={19--36},
  year={2022}
}

@inproceedings{zhang2023loopdelta,
  title={$\{$LoopDelta$\}$: Embedding Locality-aware Opportunistic Delta Compression in Inline Deduplication for Highly Efficient Data Reduction},
  author={Zhang, Yucheng and Jiang, Hong and Feng, Dan and Jiang, Nan and Qiu, Taorong and Huang, Wei},
  booktitle={2023 USENIX Annual Technical Conference (USENIX ATC 23)},
  pages={133--148},
  year={2023}
}

@INPROCEEDINGS{minervafs,
  author={Nielsen, Lars and Burihabwa, Dorian and Schiavoni, Valerio and Felber, Pascal and Lucani, Daniel E.},
  booktitle={2021 40th International Symposium on Reliable Distributed Systems (SRDS)}, 
  title={{MinervaFS: A User-Space File System for Generalised Deduplication: (Practical experience report)}}, 
  year={2021},
  volume={},
  number={},
  pages={254-264},
  doi={10.1109/SRDS53918.2021.00033}}

@ARTICLE{inlinededup_nvme,
  author={Liu, Haikun and Jin, Xiaozhong and Ye, Chencheng and Liao, Xiaofei and Jin, Hai and Zhang, Yu},
  journal={IEEE Transactions on Computers}, 
  title={{I/O Causality Based In-Line Data Deduplication for Non-Volatile Memory Enabled Storage Systems}}, 
  year={2024},
  volume={73},
  number={5},
  pages={1327-1340},
  doi={10.1109/TC.2024.3365961}}

@INPROCEEDINGS{fused_table_scan,
  author={Dreseler, Markus and Kossmann, Jan and Frohnhofen, Johannes and Uflacker, Matthias and Plattner, Hasso},
  booktitle={2018 IEEE 34th International Conference on Data Engineering Workshops (ICDEW)}, 
  title={{Fused Table Scans: Combining AVX-512 and JIT to Double the Performance of Multi-Predicate Scans}}, 
  year={2018},
  volume={},
  number={},
  pages={102-109},
  doi={10.1109/ICDEW.2018.00024}}

@INPROCEEDINGS{avx_rle,
  author={Ungethum, Annett and Pietrzyk, Johannes and Damme, Patrick and Habich, Dirk and Lehner, Wolfgang},
  booktitle={2018 IEEE 34th International Conference on Data Engineering Workshops (ICDEW)}, 
  title={{Conflict Detection-Based Run-Length Encoding - AVX-512 CD Instruction Set in Action}}, 
  year={2018},
  volume={},
  number={},
  pages={96-101},
  doi={10.1109/ICDEW.2018.00023}}

@inproceedings {sgxdedup,
author = {Yanjing Ren and Jingwei Li and Zuoru Yang and Patrick P. C. Lee and Xiaosong Zhang},
title = {{Accelerating Encrypted Deduplication via SGX}},
booktitle = {2021 USENIX Annual Technical Conference (USENIX ATC 21)},
year = {2021},
isbn = {978-1-939133-23-6},
pages = {957--971},
url = {https://www.usenix.org/conference/atc21/presentation/ren-yanjing},
publisher = {USENIX Association},
month = jul
}

@ARTICLE{securededup_video,
  author={Zheng, Yifeng and Yuan, Xingliang and Wang, Xinyu and Jiang, Jinghua and Wang, Cong and Gui, Xiaolin},
  journal={IEEE Transactions on Multimedia}, 
  title={{Toward Encrypted Cloud Media Center With Secure Deduplication}}, 
  year={2017},
  volume={19},
  number={2},
  pages={251-265},
  doi={10.1109/TMM.2016.2612760}}

@article{hashtables_vector,
author = {B\"{o}ther, Maximilian and Benson, Lawrence and Klimovic, Ana and Rabl, Tilmann},
title = {{Analyzing Vectorized Hash Tables across CPU Architectures}},
year = {2023},
issue_date = {July 2023},
publisher = {VLDB Endowment},
volume = {16},
number = {11},
issn = {2150-8097},
url = {https://doi.org/10.14778/3611479.3611485},
doi = {10.14778/3611479.3611485},
journal = {Proceedings of the VLDB Endowment},
month = jul,
pages = {2755–2768},
numpages = {14}
}

@article{clhash_vectorized,
  title={Faster 64-bit universal hashing using carry-less multiplications},
  author={Lemire, Daniel and Kaser, Owen},
  journal={Journal of Cryptographic Engineering},
  volume={6},
  pages={171--185},
  year={2016},
  publisher={Springer}
}

@inproceedings{murmur_vectorized,
  title={Optimizing high performance distributed memory parallel hash tables for {DNA} k-mer counting},
  author={Pan, Tony C and Misra, Sanchit and Aluru, Srinivas},
  booktitle={2018 International Conference for High Performance Computing, Networking, Storage and Analysis (SC)},
  pages={135--147},
  year={2018},
  organization={IEEE}
}

@techreport{sha1,
  title={{RFC 3174: US secure hash algorithm 1 (SHA1)}},
  author={Eastlake 3rd, D},
  institution={Network Working Group},
  year={2001}
}

@techreport{md5,
  title={{RFC 1321: The MD5 message-digest algorithm}},
  author={Rivest, Ronald},
  institution={Network Working Group},
  year={1992}
}

@techreport{sha256_512,
  title={{RFC 4634: US Secure Hash Algorithms (SHA and HMAC-SHA)}},
  author={Eastlake 3rd, D. and Hansen, T.},
  institution={Network Working Group},
  year={2006}
}

@article{appleby2008murmurhash,
  title={{MurmurHash3}},
  author={Appleby, Austin},
  year={2011}
}

@article{hashing_survey_xxhash,
  title={Hashing techniques: A survey and taxonomy},
  author={Chi, Lianhua and Zhu, Xingquan},
  journal={ACM Computing Surveys (Csur)},
  volume={50},
  number={1},
  pages={1--36},
  year={2017},
  publisher={ACM New York, NY, USA}
}

@article{appleby2016smhasher,
  title={{SMHasher}},
  author={Appleby, Austin},
  volume={29},
  pages={2016},
  year={2016}
}

@inproceedings{scalable_incremental_checkpt,
author = {Tan, Nigel and Luettgau, Jakob and Marquez, Jack and Teranishi, Keita and Morales, Nicolas and Bhowmick, Sanjukta and Cappello, Franck and Taufer, Michela and Nicolae, Bogdan},
title = {Scalable Incremental Checkpointing using GPU-Accelerated De-Duplication},
year = {2023},
isbn = {9798400708435},
publisher = {Association for Computing Machinery},
address = {New York, NY, USA},
url = {https://doi.org/10.1145/3605573.3605639},
doi = {10.1145/3605573.3605639},
booktitle = {Proceedings of the 52nd International Conference on Parallel Processing},
pages = {665–674},
numpages = {10},
location = {Salt Lake City, UT, USA},
series = {ICPP '23}
}

@misc{xxhashWebsite,
	author = {xxHash},
	title = {xx{H}ash - {E}xtremely fast non-cryptographic hash algorithm},
	howpublished = {\url{https://xxhash.com/}},
	year = {2020},
}

@book{altivec_instructions,
author = {Miller, Frederic P. and Vandome, Agnes F. and McBrewster, John},
title = {AltiVec},
year = {2010},
isbn = {6131814295},
publisher = {Alpha Press},
}

@inproceedings{understanding_dedup_ratios,
  title={Understanding data deduplication ratios},
  author={Dutch, Mike},
  booktitle={SNIA Data Management Forum},
  volume={7},
  year={2008}
}

@misc{rdb_format,
	author = {Rediger, Jan-Erik},
	title = {{RDB File Format}},
	howpublished = {\url{https://rdb.fnordig.de/file_format.html}},
	year = {2015},
}

@article{xxhash_weak_ssd,
title = {WOJ: Enabling Write-Once Full-data Journaling in SSDs by using weak-hashing-based deduplication},
journal = {Performance Evaluation},
volume = {127-128},
pages = {56-69},
year = {2018},
issn = {0166-5316},
doi = {https://doi.org/10.1016/j.peva.2018.09.004},
url = {https://www.sciencedirect.com/science/article/pii/S0166531618302608},
author = {Fan Ni and Xingbo Wu and Weijun Li and Lei Wang and Song Jiang},
}

@inproceedings{vectorcdc_fast,
  title={$\{$VectorCDC$\}$: Accelerating Data Deduplication with Vector Instructions},
  author={Udayashankar, Sreeharsha and Baba, Abdelrahman and Al-Kiswany, Samer},
  booktitle={23rd USENIX Conference on File and Storage Technologies (FAST 25)},
  pages={513--522},
  year={2025}
}

\end{document}